\documentclass[reprint,amsmath,amssymb, notitlepage,prb,superscriptaddress,floatfix]{revtex4-2}
\usepackage{graphicx}
\usepackage{dcolumn}
\usepackage{bm}
\usepackage{physics}
\usepackage{color}
\usepackage{comment}

\allowdisplaybreaks[1]

\usepackage[normalem]{ulem}

\usepackage[whole]{bxcjkjatype} 

\newcommand{\av}[1]{\ensuremath{\langle#1\rangle} }
\newcommand{\ave}[1]{\ensuremath{\left\langle#1\right\rangle} }

\DeclareMathOperator{\sgn}{sgn}

\usepackage{hyperref}

\begin{document}
	
\title{Microscopic theory of spin-torque ferromagnetic resonance \\
in nonmagnetic-metal/ferromagnetic-metal heterostructures}

\author{Takumi Funato}
\affiliation{Advanced Science Research Center, Japan Atomic Energy Agency, Tokai, 319-1195, Japan}

\author{Takeo Kato}
\affiliation{Institute for Solid State Physics, University of Tokyo, Kashiwa, 277-8581, Japan}
	
\date{\today}

\begin{abstract}
We develop a microscopic theory of spin-torque ferromagnetic resonance (ST-FMR) in nonmagnetic-metal/ferromagnetic-metal heterostructures.  
Tracing out the conduction-electron degrees of freedom in the FM, we derive an effective interfacial exchange coupling between the localized spins and the conduction electron spins in the adjacent electron system.
Based on this interaction, we calculate the current-induced driving torque, resonance-frequency shift, damping modulation, and resulting dc voltage.
As a concrete example, we apply the formulation to a disordered Rashba two-dimensional electron gas and demonstrate that the ST-FMR spectrum reflects the dynamical spin responses of the adjacent electron system. 
Our formulation applies to a broad class of heterostructures and establishes a unified microscopic framework connecting ST-FMR spectra directly to the electronic spin responses of adjacent systems.
\end{abstract}

\maketitle

\section{Introduction}

Spin-torque ferromagnetic resonance (ST-FMR) has become a standard experimental tool for investigating current-induced magnetization dynamics in magnetic heterostructures. 
The electrical detection of resonant magnetization precession through rectification of an rf current was demonstrated in pioneering studies on magnetic tunnel junctions and metallic nanomagnets, establishing ST-FMR as a powerful probe of spin-transfer torque in nanoscale systems~\cite{Tulapurkar2005,kubota2008,sankey2008,Sankey2006}.
The method was later extended to bilayer systems in which spin-orbit coupling in the nonmagnetic layer generates the driving torque, establishing ST-FMR as a practical tool for quantifying spin-orbit torques~\cite{Liu2011,pai2012,pai2015a,he2016,berger2018}.
Beyond simple metallic bilayers, ST-FMR is now widely applied to systems where the band structure and spin texture of the adjacent conductor strongly influence the generated torque, including interfaces with spin-orbit coupling (SOC), two-dimensional electron systems, and topological materials\cite{skinner2015,chen2016,jungfleisch2016,karube2020,wang2017,macneill2017,mellnik2014,kondou2016}.
The scope of ST-FMR has further expanded to alternative torque-generation mechanisms, such as orbital torque and current-vorticity-induced spin torque\cite{lee2021,hayashi2023, nakayama2023,horaguchi2025}.

\begin{figure}[b]
\centering
\includegraphics[width=75mm]{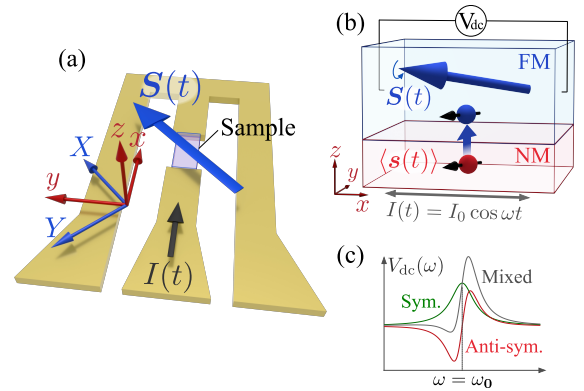}
\caption{
Schematic illustration of the ST-FMR setup and dc-voltage spectrum.
(a) Coplanar-waveguide geometry with the NM/FM sample and localized-spin direction.
(b) Current-induced spin accumulation near the NM/FM interface drives the magnetization dynamics.
(c) Representative dc-voltage spectrum consisting of symmetric and antisymmetric components.
}
\label{fig:setup}
\end{figure}

The interpretation of ST-FMR experiments has commonly been based on phenomenological magnetization dynamics combined with interfacial spin-transport theory. 
To describe the spin-orbit torque driving the magnetization, spin transport in the adjacent conductor is commonly treated within drift-diffusion theory, while interfacial spin transfer is characterized by the spin-mixing conductance within magnetoelectronic circuit theory~\cite{Brataas2000,Tserkovnyak2002,Tserkovnyak2005,Zwierzycki2005}. 
The resulting driving torque is incorporated into the Landau–Lifshitz–Gilbert equation to describe the magnetization dynamics, which, together with magnetoresistive rectification, determines the dc voltage generated by ST-FMR (see Fig.~\ref{fig:setup}).
This theoretical framework provides a description of the resonance-frequency, magnetization-direction, and layer-thickness dependences of the dc voltage~\cite{Liu2011,Chiba2014,Sklenar2015,Karimeddiny2020}.
These theoretical descriptions are widely used in the analysis of ST-FMR experiments to evaluate the magnitude, direction, and symmetry of spin-orbit torques.

Despite its practical success, this conventional interpretation has an important limitation.
In many cases, the spin-mixing conductance is treated as an effective interfacial parameter rather than being derived microscopically from the electronic structure and spin dynamics of the adjacent conductor.
It is therefore difficult to predict its temperature dependence, material dependence, and sensitivity to disorder and spin-orbit interaction on equal footing.
Recent microscopic studies of spin pumping have demonstrated that modifications of the FMR signal can reflect the spin excitations and electronic structures of adjacent electron systems~\cite{ohnuma2014,kato2019,yama2021,yama2023,Ominato2025}. Similarly, current-induced spin torques originate from the nonequilibrium spin response of the adjacent electron system and therefore depend sensitively on its electronic structure, spin-orbit interaction, and scattering processes~\cite{Haney2013,Manchon2019,Zhu2019}. 
These developments suggest that ST-FMR can provide information about both the current-induced spin response and the dynamical spin susceptibility of adjacent electron systems. In ST-FMR, the former determines the driving torque and its effective magnetic field, whereas the latter governs the resonance-frequency shift and damping modulation. 
The ST-FMR dc voltage reflects the combined effects of these quantities through the resonant magnetization dynamics. A microscopic framework that consistently connects these electronic spin responses to the ST-FMR voltage is therefore desirable.

The purpose of this work is to provide a microscopic theoretical framework by formulating ST-FMR within a Green-function-based theory. We consider a heterostructure composed of a nonmagnetic metal (NM) and a ferromagnetic metal (FM).
We first express the dc voltage generated by ST-FMR in terms of the retarded magnon Green function and the effective magnetic field associated with the spin-orbit torque. 
To microscopically connect the localized spins in the FM to the itinerant spins in the NM, we integrate out the FM conduction electrons and derive an effective interfacial exchange action. 
Using this interaction, we derive both the driving torque governed by the current-induced spin density in the NM and the magnon self-energy determined by the dynamical spin susceptibility, with the latter giving rise to the FMR frequency shift and damping modulation.
As a concrete example, we apply the formulation to a two-dimensional electron gas (2DEG) with Rashba SOC.

\section{Outline}
\label{sec:setup}

We consider an ST-FMR setup consisting of a bilayer of a nonmagnetic metal (NM) and a ferromagnetic metal (FM), as schematically illustrated in Fig.~\ref{fig:setup}.
We introduce the laboratory coordinates so that the $x$ axis is set along the current through the target (the red part), and the $z$ axis is directed normal to the junction interface.
We denote the localized spin in the FM per site with ${\bm S}$.
We assume that a stable spin direction in the FM is fixed as $\langle {\bm S} \rangle = (S \cos \theta, S \sin \theta, 0)$ in the laboratory coordinates by an in-plane external magnetic field, where $S$ is the amplitude of the spin per site.

By applying ac charge current $I(t) = I_0 \cos(\omega t)$ to the bilayer system, the spin dynamics in the FM is excited by spin-orbit torque, which originates from spin accumulation in the adjacent NM due to the spin Hall effect. 
We note that the Oersted field also affects the spin dynamics in the FM.
Since anisotropic magnetoresistance (AMR) in the FM is modulated by spin precession, the total resistance of the heterostructure is periodically modulated as
\begin{align}
R(t) = R_0 + R_A \frac{\av{S^x(t)}^2}{S^2}, \label{eq:r}
\end{align}
where $R_0$ and $R_A$ are the sample-dependent parameters with a dimension of resistance.
We note that the AMR contribution depends only on the spin component parallel to the current.
By combining this oscillating resistance with the external ac current, the time-dependent voltage $V(t) = R(t) I(t)$ across the bilayer sample includes a net dc voltage, whose amplitude is given by taking the time average over one period $T=2\pi /\omega$ as
\begin{align}
        V_{\text{dc}} &= \frac{R_A}{T} \int^T_0  \frac{\av{S^x(t)}^2}{S^2} I(t) dt.
        \label{eq:V_dc}
\end{align}
Thus, the dc voltage measured in the ST-FMR experiment reflects nonequilibrium spin dynamics of the FM through $\av{S^x(t)}$.

\begin{figure}[tb]
\centering
\includegraphics[width=70mm]{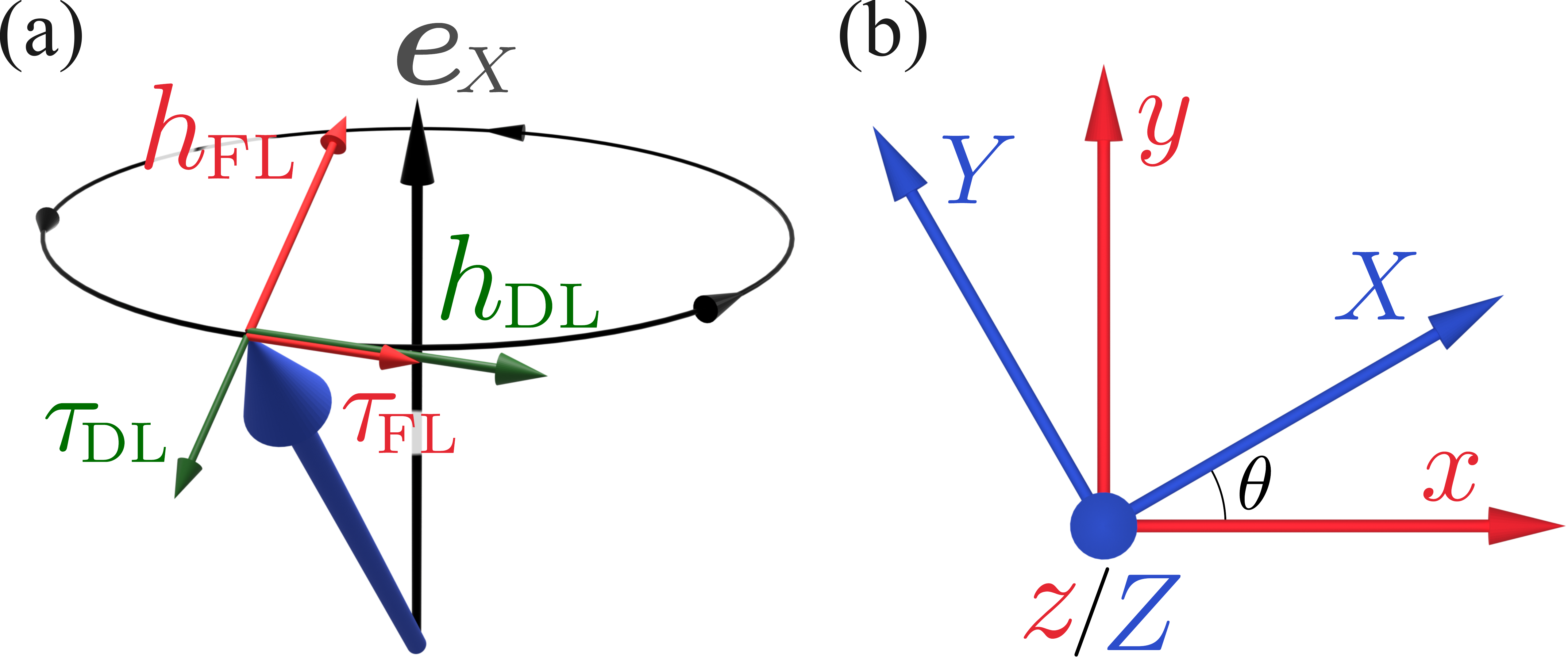}
\caption{
Schematic illustration of (a) the localized-spin dynamics driven by the damping-like $\tau_{\rm DL}$ and field-like $\tau_{\rm FL}$ torques and (b) the coordinate systems.
The corresponding effective magnetic fields are denoted by $\bm h_{\rm DL}$ and $\bm h_{\rm FL}$, respectively.
The $X$-axis of the magnetization-fixed coordinate system $(X,Y,Z)$ is tilted from the $x$-axis of the original coordinate system $(x,y,z)$ by an angle $\theta$.
}
\label{fig:fig2}
\end{figure}

To describe spin dynamics in the FM, it is convenient to introduce the magnetization-fixed coordinates \((X,Y,Z)\).
The \(X\)-axis is chosen to be parallel to the equilibrium direction of the localized spin keeping the $Z$-axis the same as the $z$ axis  (see Fig.~\ref{fig:fig2}).
In the new coordinates, the average $\av{S^x(t)}$ is described  as
\begin{align}
    \av{S^x(t)} &= \cos \theta \av{S^X(t)} - \sin \theta \av{S^Y(t)} \notag \\
    &\simeq S \cos \theta - \sin \theta \av{S^Y(t)} ,
    \label{eq:S_x}
\end{align}
where the variation of the spin direction from the equilibrium state is assumed to be small.

The effect of current-induced spin-orbit torque due to the adjacent NM can be represented by an effective Hamiltonian
\begin{align}
\mathcal{V}(t)
 &= - N_{\rm FM} \hbar\gamma_g {\bm h}(t)  \cdot {\bm S},
\label{eq:Zeeman_perturbation} 
\end{align}
where $N_{\rm FM}$ is the number of spins in the FM,  $\gamma_g$ is the gyromagnetic ratio of the FM, ${\bm S}=N_{\rm FM}^{-1} \sum_j {\bm S}_j$ is the spin operator per site, and ${\bm h}(t) = (h_X(t),h_Y(t),h_Z(t))$ is an effective Zeeman field acting on the spin at the FM interface.
Hereafter, we omit the longitudinal component $h_X(t)$, which is parallel to the equilibrium magnetization, and retain only the transverse components that exert a torque to linear order.
The torque is commonly decomposed into damping-like and field-like components, whose geometrical relation to the corresponding effective fields is schematically illustrated in Fig.~\ref{fig:fig2}.
The transverse components of the driving field contribute to the spin dynamics within the linear-response regime. 
Introducing
$S^\pm = S^Y \pm iS^Z$ and $h_\pm(t) = h_Y(t) \pm i h_Z(t)$ with $h_{Y,Z}(t)=\Re[h_{Y,Z}(\omega)e^{-i\omega t}]$, the transverse part of Eq.~\eqref{eq:Zeeman_perturbation} can be rewritten as
\begin{align}
\mathcal{V}(t)
 = -\frac{N_{\rm FM}\hbar\gamma_g}{2}
 \left[
 h_-(t) S^+ + h_+(t) S^-
 \right].
\label{eq:Zeeman_transverse}
\end{align}
Introducing the complex amplitudes $h_\pm(\omega) \equiv h_Y(\omega) \pm i h_Z(\omega)$,
the field coupled to $S^-$ is written as
\begin{align}
h_+(t) = \frac{1}{2}\left[ h_+(\omega) e^{-i\omega t} + h_-(\omega)^* e^{i\omega t} \right].
\end{align}
Near the resonance, $\omega \simeq \omega_{\bm 0}$, the counter-rotating component
$\propto e^{i\omega t}$ yields only a nonresonant contribution of relative order
$\alpha_{\rm G}$ and is neglected hereafter (rotating-wave approximation).

By the linear response theory, the spin dynamics is described as
\begin{align}
\delta \langle S^+(t)\rangle &= \delta \langle S^+(\omega) \rangle e^{-i\omega t}, \\
\delta\langle S^+(\omega)\rangle&=-\dfrac{\hbar\gamma_g}{4}G^R(\omega)h_+(\omega), 
\label{eq:Splus_response} \\
G^{R}(\omega)
 &= -\frac{iN_{\rm FM}}{\hbar}
 \int_0^\infty dt\,
 e^{i(\omega+i\delta)t}
 \left\langle
 \left[
 S^{+}(t),S^{-}(0)
 \right]
 \right\rangle .
\label{eq:magnon_GF_definition}
\end{align}
The linear response of the $Y$ component is then given by 
\begin{align}
\delta\langle S^Y(t)\rangle=-\dfrac{\hbar\gamma_g}{4}\Re[G^R(\omega)h_+(\omega)e^{-i\omega t}].
\label{eq:S_Y}
\end{align}
Substituting the nonequilibrium spin response obtained in Eqs.~(\ref{eq:S_x}) and (\ref{eq:S_Y}) into the AMR in Eq.~(\ref{eq:V_dc}), we obtain the dc voltage as
\begin{align}
    V_{\rm dc} &= \frac{R_A I_0 \hbar \gamma_g}{8S} \sin 2\theta \notag \\
    &\times 
    \Bigl( \Re[G^R(\omega) h_Y(\omega)] - \Im[G^R(\omega) h_Z(\omega)] \Bigr).
    \label{eq:dc_voltage}
\end{align}
We note that only the in-phase component of $\delta \av{S^Y (t)}$ with respect to the current $I(t) = I_0 \cos (\omega t)$ gives a finite contribution after time averaging.

Thus, to evaluate the dc voltage in ST-FMR, we need to calculate two ingredients, i.e., the effective magnetic fields ($h_Y(\omega)$ and $h_Z(\omega)$) and the response function $G^R(\omega)$.
The main purpose of this work is to formulate these ingredients from a microscopic model.
In Sec.~\ref{sec:eff}, we introduce a microscopic model for the NM/FM junction and provide a general formalism for perturbation theory with respect to the interfacial coupling.
Subsequently, we develop a microscopic formulation of $h_Y(\omega)$, $h_Z(\omega)$, and $G^R(\omega)$ in Sec.~\ref{sec:effective_magnetic_field}.

\section{Microscopic description}\label{sec:eff}

\subsection{Model}

\begin{figure}
\centering
\includegraphics[width=45mm]{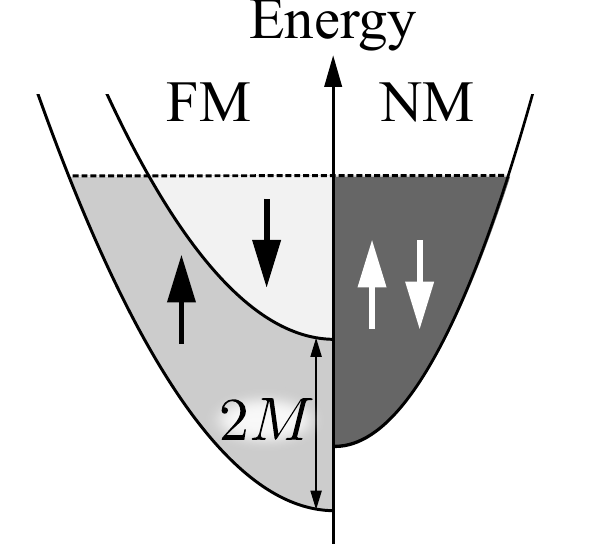}
\caption{Schematic illustration of the band structures of the FM and the NM. The conduction electron energy is spin-split by the $s$-$d$ exchange coupling.}
\label{fig:dispersion1}
\end{figure}

We consider a bilayer system composed of the FM and the NM for which the Hamiltonian is given by
\begin{align}   
\hat{\mathcal H} = \hat{\mathcal H}_{\rm NM} + \hat{\mathcal H}_{\rm FM} + \hat{\mathcal H}_t,
\label{eq:H_tot}
\end{align}
where $\hat{\mathcal H}_{\rm NM}$ and $\hat{\mathcal H}_{\rm FM}$ respectively describe the NM and FM, and $\hat{\mathcal H}_t$ represents an interfacial coupling.
In the following, we explicitly introduce the Hamiltonians $\hat{\mathcal H}_{\rm FM}$ and $\hat{\mathcal H}_t$, while the Hamiltonian for the NM $\hat{\mathcal H}_{\rm NM}$ is kept arbitrary in our general formulation given in this section.

The Hamiltonian for the FM is described by the $s$-$d$ model 
\begin{align}
\hat{\mathcal H}_{\rm FM} &= \hat{\mathcal H}_s + \hat{\mathcal H}_d +  \hat{\mathcal H}_{s\text{-}d}. 
\end{align}
Here, $\hat{\mathcal H}_s$ describes the conduction electrons in the FM, $\hat{\mathcal H}_d$ describes the localized spin, and $\hat{\mathcal H}_{s\text{-}d}$ denotes the exchange coupling between the conduction electrons and the localized spin.
The conduction electrons are described by noninteracting electrons as
\begin{align}
\hat{\mathcal H}_{s} = \sum_{\bm l\sigma} \xi^{\rm F}_{\bm l\sigma} a^\dagger_{\bm l\sigma} a_{\bm l\sigma},
\end{align}
where $a^\dagger_{\bm l\sigma}$ ($a_{\bm l\sigma}$) is the creation (annihilation) operator of conduction electrons with wavenumber ${\bm l}$ and spin $\sigma$ in the FM, $\xi^{\rm F}_{\bm l\sigma} = \hbar^2 {\bm l}^2/2m - \mu - \sigma M$ is the spin-split energy measured from the chemical potential $\mu$, and $2M$ is the exchange splitting of the conduction electrons in the FM (see Fig.~\ref{fig:dispersion1}).
The localized spins in the FM are described by the Heisenberg model
\begin{align}
\hat{\mathcal H}_d = - J\sum_{\av{i,j}} \bm S_i \cdot \bm S_j - \hbar \gamma_g h_{\text{dc}} \sum_i S^X_i, 
\end{align}
where $\av{i,j}$ denotes a pair of nearest-neighbor sites, $J$ is the ferromagnetic exchange coupling, and $h_{\text{dc}}$ is the magnitude of the static magnetic field.
We assume that the magnetization is aligned in the $X$ direction.
To consider a small transverse fluctuation around the $X$ direction at low temperatures, we employ the spin-wave approximation. We define the Fourier transformation of the spin operators as \begin{align} S_{\bm q}^\alpha = \frac{1}{\sqrt{N_{\rm FM}}} \sum_{j} e^{-i{\bm q}\cdot {\bm r}_j} S_{j}^\alpha, \quad (\alpha = X,Y,Z,+,-), \end{align}
where $N_{\rm FM}$ is the number of unit cells in the FM, and ${\bm r}_j$ is the position of the site $j$.
Using the Holstein-Primakoff transformation, the localized spin operators are written as
\begin{align}
S^X_{\bm q} &= \sqrt{N_{\rm FM}} S \delta_{\bm q,\bm 0}  - \frac{1}{\sqrt{N_{\rm FM}}}\sum_{\bm k} b_{\bm k}^\dagger b_{\bm k+\bm q}, \\
S^+_{\bm q} &= S^Y_{\bm q} + iS^Z_{\bm q} = \sqrt{2S} b_{\bm q} + {\cal O}(S^{-1/2}), \\
S^-_{-{\bm q}} & = (S^+_{\bm q})^\dagger = \sqrt{2S} b_{\bm q}^\dagger + {\cal O}(S^{-1/2}),
\end{align}
where $b_{\bm q}^\dagger$ ($b_{\bm q}$) is a magnon creation (annihilation) operator.
Then, the Hamiltonian for the localized spin is rewritten in the leading order of $1/S$ as
\begin{align}
\hat{\mathcal H}_d &= \sum_{\bm q} \hbar \omega_{\bm q} b^\dagger_{\bm q} b_{\bm q},
\end{align}
where $\hbar \omega_{\bm q} = \mathcal D q^2 + \hbar \gamma_g h_{\text{dc}}$ with $\mathcal D$ being the spin stiffness.
The $s$-$d$ exchange interaction $\hat{\mathcal H}_{{s\text{-}d}}$ is given by 
\begin{align}
\hat{\mathcal H}_{s\text{-}d}&=-J_{s\text{-}d}\sum_j\bm S_j\cdot\hat{\bm s}_j, \\
\hat{\bm s}_j &=\sum_{\sigma\sigma'} a_{j\sigma}^\dagger  \left(\frac{\bm\sigma}{2}\right)_{\sigma \sigma'} a_{j\sigma'}, \\
a_{j\sigma} &= \frac{1}{\sqrt{N_{\rm FM}}} \sum_{\bm l} e^{i{\bm l}\cdot {\bm r}_j} a_{{\bm l}\sigma}, 
\end{align}
where $J_{s\text{-}d}$ represents the strength of the $s$-$d$ exchange interaction, $\bm \sigma = (\sigma^X, \sigma^Y, \sigma^Z)$ are the Pauli matrices, and $\hat{\bm s}_j$ is the spin operator of the conduction electrons in the FM.
Performing the Fourier transformation, the $s$-$d$ exchange interaction $\hat{\mathcal H}_{{s\text{-}d}}$ is rewritten as
\begin{align}
\hat{\mathcal H}_{{s\text{-}d}} = -\frac{J_{s\text{-}d}}{2\sqrt{N_{\rm FM}}}\sum_{\bm l\bm q\sigma \sigma'} a^\dagger_{\bm l+\bm q\sigma} (\bm \sigma)_{\sigma \sigma'} a_{\bm l\sigma'}  \cdot \bm S_{\bm q}.
\label{eq:H_sd}
\end{align}
In the leading order of $1/S$, the exchange interaction for the $X$ component gives a static Zeeman field on the conduction electrons, and can be incorporated into the spin-split energy $M = J_{s\text{-}d} S/2$.
For the other transverse components, the exchange interaction is written as
\begin{align}
\hat{\mathcal H}_{{s\text{-}d}} &= -\frac{J_{s\text{-}d}}{2\sqrt{N_{\rm FM}}} \sum_{\bm l\bm q} (a^\dagger_{\bm l+\bm q\uparrow} a_{\bm l\downarrow} S^-_{\bm q} + 
a^\dagger_{\bm l+\bm q \downarrow}  a_{\bm l\uparrow} S^+_{\bm q} ).
\label{eq:fluc_sd}
\end{align}

The Hamiltonian for electron tunneling through the interface of the NM/FM junction is given by
\begin{align}
\hat{\mathcal H}_t = \sum_{\bm k\bm l\sigma} \qty( T^\sigma_{\bm l,\bm k}  a^\dagger_{\bm l\sigma} c_{\bm k\sigma} + T^{\sigma *}_{\bm l,\bm k}  c^\dagger_{\bm k\sigma} a_{\bm l\sigma}),
\label{eq:tunnel}
\end{align}
where $c^\dagger_{\bm k\sigma} (c_{\bm k\sigma})$ is the creation (annihilation) operator of conduction electrons in the NM.
For simplicity, we assume that the tunneling matrix elements $T^\sigma_{\bm l,\bm k}$ are independent of the spin.

\subsection{Effective interaction}
\label{sec:effective_interaction}

\begin{figure}
    \centering
    \includegraphics[width=0.85\linewidth]{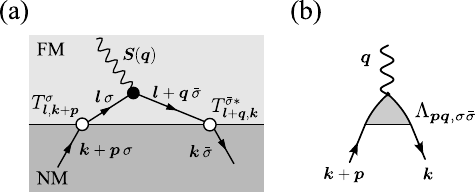}
    \caption{(a) Schematic illustration of the interfacial exchange process involving one $s$-$d$ exchange interaction and two interfacial electron-tunneling processes. (b) Corresponding Feynman diagram.}
    \label{fig:diagram1}
\end{figure}

In our model, the localized spins in the FM couple to the FM conduction electrons through the $s$-$d$ exchange interaction, whereas the FM and NM conduction electrons are connected by interfacial electron tunneling. By integrating out the FM conduction-electron degrees of freedom, these microscopic processes generate an effective coupling between the localized spins in the FM and the conduction-electron spins in the NM. 
We take the unperturbed Hamiltonian as
$\hat{\mathcal H}_0=\hat{\mathcal H}_s+\hat{\mathcal H}_d+\hat{\mathcal H}_{\rm NM}$
and define the dimensionless effective interaction action by
\begin{align}
e^{-\mathcal S_{\rm int}}  =
\left\langle
T_\tau
\exp\left[
- \int_0^{\beta}d\tau\,
\left\{
\hat H_{s\text{-}d}(\tau)+\hat H_t(\tau)
\right\}
\right]
\right\rangle_s .
\label{eq:Seff_definition}
\end{align}
Here,
$\hat O(\tau)=e^{\tau\hat{\mathcal H}_0}\hat Oe^{-\tau\hat{\mathcal H}_0}$,
and $\langle\cdots\rangle_s$ denote the partial trace over the FM conduction-electron degrees of freedom.
The lowest-order process that directly couples the NM electron spin to the localized FM spin consists of one $s$-$d$ exchange interaction and two tunneling processes, as shown in Fig.~\ref{fig:diagram1}. 
The corresponding third-order contribution is
\begin{align}
    \mathcal S_{\rm int}
&= \frac{1}{2! }
\int_0^{\beta}d\tau_1\,d\tau_2\,d\tau_3\,
\notag \\
&\qquad \qquad
\times \left\langle
T_\tau
\hat H_{s\text{-}d}(\tau_1)
\hat H_t(\tau_2)
\hat H_t(\tau_3)
\right\rangle_{s,c},
\label{eq:Seff_third_order}
\end{align}
where the subscript $c$ denotes the connected contribution.
We note that the contribution of order  $J_{s\text{-}d}^{2}T^{2}$ gives the leading contribution to the magnon self-energy in the tunneling amplitude.
Within the flat wide-band approximation, however, it depends only on the NM density of states at the Fermi level and does not reflect its spin structure; we therefore do not consider it further in the main text (detailed derivation shown in Appendix~\ref{appsec:spin_pump}).

Here, we introduce the average over the interfacial randomness denoted with $\langle \cdots \rangle_{\rm av}$, and apply it to the lowest-order contribution, Eq.~\eqref{eq:Seff_third_order}.
We define the random-averaged tunneling correlation as $\mathcal T_{\bm p,\bm q}=\langle T^\sigma_{\bm l,\bm k+\bm p}T^{\bar\sigma *}_{\bm l+\bm q,\bm k}\rangle_{\rm av}$, where it is assumed to be independent of ${\bm k}$ and ${\bm l}$ for simplicity.
Applying Wick's theorem to the FM conduction-electron operators and Fourier transforming to Matsubara frequencies, the effective action can be written as
\begin{align}
\mathcal S_{\rm int}
=
-&
\sum_{\bm p\bm q, n\lambda}
\Bigl[
\Lambda_{\bm p\bm q,\uparrow\downarrow}(i\varepsilon_n,i\omega_\lambda)
S_{\bm q}^{+}(i\omega_\lambda)
s_{\bm p}^{-}(i\varepsilon_n,i\omega_\lambda)
\nonumber\\
+{}&
\Lambda_{\bm p\bm q,\downarrow\uparrow}(i\varepsilon_n,i\omega_\lambda)
S_{\bm q}^{-}(i\omega_\lambda)
s_{\bm p}^{+}(i\varepsilon_n,i\omega_\lambda)
\Bigr],
\label{eq:effective_action}
\end{align}
where the energy-resolved operators are defined by
\begin{align}
c_{\bm k\sigma}(i\varepsilon_n) &=
\frac{1}{\sqrt{\beta}} \int_0^{ \beta} d\tau \, e^{i\varepsilon_n\tau }
c_{\bm k\sigma}(\tau), \\
S_{\bm q}^\pm (i\omega_\lambda) &=
\frac{1}{\beta} \int_0^{ \beta} d\tau \, e^{i\omega_\lambda \tau }
S^\pm_{\bm q}(\tau), \\
s_{\bm p}^{\pm}(i\varepsilon_n,i\omega_\lambda)
&=
\frac{1}{2}
\sum_{\bm k,\sigma}
c_{\bm k\bar \sigma}^{\dagger}(i\varepsilon_n+i\omega_\lambda)
\sigma^\pm_{\bar \sigma \sigma}
c_{\bm k+\bm p \sigma}(i\varepsilon_n),
\label{eq:NM_spin_density_energy} 
\end{align}
where $\sigma^\pm = \sigma^Y \pm i \sigma^Z$, $\varepsilon_n = (2n+1)\pi/\beta$ are the fermion Matsubara energies, and $\omega_\lambda = 2\pi \lambda/\beta$ are the bosonic Matsubara energies.
The effective interfacial vertex is given by
\begin{align}
& \Lambda_{\bm p\bm q,\sigma\bar\sigma}(i\varepsilon_n,i\omega_\lambda) \nonumber \\
&=
\frac{J_{s\text{-}d}}{2\sqrt{N_{\rm FM}}}
\mathcal T_{\bm p,\bm q}
 \sum_{\bm l}
D_{\bm l\sigma}(i\varepsilon_n)
D_{\bm l+\bm q\bar\sigma}(i\varepsilon_n+i\omega_\lambda),
\label{eq:FM_kernel}
\end{align}
where we define the Green function of the FM conduction electrons as
\begin{align}
D_{\bm l\sigma}(i\varepsilon_n)
=
\frac{1}
{i\varepsilon_n- \xi_{\bm l} + \sigma M +i\gamma_\sigma\sgn(\varepsilon_n)},
\label{eq:FM_conduction_green-func}
\end{align}
with $\xi_{\bm l}=\hbar^2 {\bm l}^2/2m - \mu$ and $\gamma_\sigma$ being the damping rate for the $\sigma$-spin conduction electrons.

We next evaluate the FM conduction-electron kernel within the flat wide-band approximation,
\begin{align}
\frac{1}{N_{\rm FM}}\sum_{\bm l}(\cdots)
\rightarrow
\tilde{\nu}_{\rm F}\int_{-\infty}^{\infty}d\xi_{\bm l}
\int\frac{d\Omega_{\bm l}}{4\pi}\,(\cdots) ,
\label{eq:sum_to_int}
\end{align}
where $\tilde{\nu}_{\rm F}$ is the density of states per site per spin at the Fermi level for the conduction electrons in the FM.
For $\omega_\lambda>0$, the kernel is finite only for $-\omega_\lambda<\varepsilon_n<0$, where the two FM electron propagators have poles in opposite half-planes.
Therefore, the vertex is written in the form
\begin{align}
\Lambda_{\bm p\bm q,\sigma\bar\sigma}(i\varepsilon_n,i\omega_\lambda) &=
\Theta(\varepsilon_n, \omega_\lambda) \widetilde{\Lambda}_{\bm p\bm q,\sigma\bar\sigma}(i\omega_\lambda), \label{eq:lambda_factorization} \\
\Theta(\varepsilon_n, \omega_\lambda) &= \theta(-\varepsilon_n)\theta(\varepsilon_n + \omega_\lambda) .
\end{align}
Within this approximation, the $\theta$-function restriction in Eq.~\eqref{eq:lambda_factorization} selects electron--hole excitations across the Fermi level, so that only the Fermi-surface contribution of the adjacent electron system is retained.
The factorized effective interfacial vertex $\widetilde{\Lambda}_{\bm p\bm q,\sigma\bar\sigma}(i\omega_\lambda)$ is calculated near the Fermi surface as
\begin{align}
\widetilde{\Lambda}_{\bm p\bm q,\sigma\bar\sigma}
(i\omega_\lambda) &= \frac{J_{s\text{-}d}\sqrt{N_{\rm FM}}}{2}\mathcal T_{\bm p,\bm q} \frac{\pi i\tilde{\nu}_{\rm F}}{\hbar v_{\rm F}q} \notag \\
&\times \ln \left[ \frac{i\omega_\lambda-2\sigma M-\varepsilon_q+i\Gamma_s+\hbar v_{\rm F}q}{i\omega_\lambda-2\sigma M-\varepsilon_q+i\Gamma_s-\hbar v_{\rm F}q}
\right].
\label{eq:lambda_finite_q}
\end{align}
Here, $\Gamma_s=\gamma_\uparrow + \gamma_\downarrow$ is the transverse spin broadening in the FM, and we have used 
$\xi_{\bm l+\bm q}-\xi_{\bm l}\simeq \hbar v_{\rm F} \hat{\bm l} \cdot \bm q + \varepsilon_q$ with $\hat{\bm l} = \bm l/l$, $v_F=\hbar l_F/m$, and $\varepsilon_q = \hbar^2q^2/2m$.

For the real-time response, analytically continuing the frequency-dependent factor as $i\omega_\lambda\rightarrow\hbar\omega+i0^+$ and taking the uniform-mode limit $\bm q=\bm 0$, relevant to the spatially uniform FMR excitation considered here, gives
\begin{align}
\widetilde{\Lambda}_{\bm p\bm0,\sigma\bar\sigma}^{R}(\omega)
=
\frac{J_{s\text{-}d}\sqrt{N_{\rm FM}}}{2}
\mathcal T_{\bm p,\bm0}
\frac{2\pi i\tilde{\nu}_{\rm F}}
{\hbar\omega-2\sigma M+i\Gamma_s},
\label{eq:Lambda_retarded}
\end{align}
which satisfies
$\widetilde{\Lambda}_{\bm p\bm0,\uparrow\downarrow}^{R}(\omega)
=
[\widetilde{\Lambda}_{-\bm p\bm0,\downarrow\uparrow}^{R}(-\omega)]^*$, where $\mathcal T_{-\bm p,\bm0}=\mathcal T_{\bm p,\bm0}^{*}$ has been used.
In the following, we focus on the low-frequency response ($\hbar\omega \ll M$), which is usually fulfilled in standard ST-FMR experimental setups.
Then, the vertex is rewritten as
\begin{align}
\widetilde{\Lambda}_{\bm p \bm 0,\sigma\bar\sigma}^{R}(\omega)
& \simeq 
-i \sigma \sqrt{N_{\rm FM}} \mathcal T_{\bm p,\bm0} \Lambda_\sigma ,
\label{eq:Lambda_retarded1} \\
\Lambda_\sigma 
&= \frac{\pi \tilde{\nu}_{\rm F}}{S} \frac{ 2M}
{2M - i\sigma\Gamma_{s}} ,
\label{eq:Lambda_retarded2}
\end{align}
where we use $J_{s\text{-}d}=2M/S$. 
Note that $\Lambda_{\uparrow} = \Lambda_{\downarrow}^*$.
Under the condition $\Gamma_s \ll 2M$, which is also a standard condition in ST-FMR, the imaginary part is much smaller than the real part ($|{\rm Im} \, \Lambda_\sigma | \ll {\rm Re} \, \Lambda_\sigma$).
In this limit, $\Lambda_\sigma$ becomes approximately real, $\Lambda_\sigma \simeq \pi\tilde{\nu}_F/S$.
Consequently, the full effective interfacial vertex $\widetilde{\Lambda}_{\bm p\bm0,\sigma\bar\sigma}^{R}$ is predominantly imaginary, reflecting the $\pi/2$ phase shift associated with the precession of the transverse spin around the exchange field of the FM conduction electrons (see Appendix~\ref{appsec:eff}).

\subsection{Effective interaction on magnons}
\label{sec:magnon_response}

\begin{figure}
    \centering
    \includegraphics[width=0.85\linewidth]{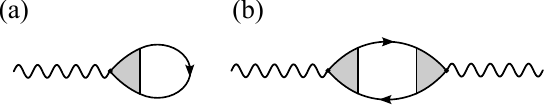}
    \caption{Feynman diagrams for the self-energy of the magnon: (a) the effective Zeeman field and (b) the FMR frequency shift and damping modulation.}
    \label{fig:diagram3}
\end{figure}

We have derived the effective interfacial interaction between the magnon in the FM and the conduction electrons of the NM by tracing out the degrees of freedom of the conduction electrons in the FM.
By further tracing out the degrees of freedom of the electrons in the NM, we define the effective action for the magnon $S_{\rm int}^{\rm mag}$ as
\begin{align}
e^{-S_{\rm int}^{\rm mag}} &=
\left\langle
e^{-\mathcal S_{\rm int}}
\right\rangle_{\rm NM} , 
\end{align}
where $\langle\cdots\rangle_{\rm NM}$ denotes the average over the NM conduction-electron degrees of freedom.

Expanding the effective interfacial interaction gives
\begin{align}
\mathcal S_{\rm int}^{\rm mag}
=
\left\langle \mathcal S_{\rm int}\right\rangle_{\rm NM}
-
\frac{1}{2}
\left\langle \mathcal S_{\rm int}^{2}\right\rangle_{{\rm NM},c}
+\mathcal O(\mathcal S_{\rm int}^{3}).
\label{eq:magnon_cumulant}
\end{align}
The first cumulant, evaluated to linear order in the applied ac electric field, generates a term linear in the localized spin and therefore acts as an effective magnetic field.
The corresponding Feynman diagram is shown in Fig.~\ref{fig:diagram3}(a) (see also Eq.~\eqref{eq:first_cumulant}).
On the other hand, the second cumulant, involving the equilibrium connected spin correlation of the NM, generates bilinear terms in the localized spin operators and renormalizes the magnon propagator.
The corresponding Feynman diagram is shown in Fig.~\ref{fig:diagram3}(b) (see also Eq.~\eqref{eq:second_cumulant}). In the subsequent two subsections, we explicitly calculate these two cumulants.

\subsection{Effective magnetic field}
\label{sec:effective_magnetic_field}

We first consider the first cumulant in Eq.~\eqref{eq:magnon_cumulant}. Since the ac electric field applied in ST-FMR is spatially uniform, we focus on the uniform current-induced spin response in the NM, $\bm p=\bm0$.
The first cumulant is then written for $\omega_\lambda>0$ as
\begin{align}
\left\langle \mathcal S_{\rm int}\right\rangle_{\rm NM}
= - \beta \sum_{\bm q,\lambda} \Bigl[
&\widetilde{\Lambda}_{\bm 0 \bm q,\uparrow\downarrow}(i\omega_\lambda)
S_{\bm q}^{+}(i\omega_\lambda)
\av{s_{\bm0}^{-}(i\omega_\lambda)}_{\rm FS}
\nonumber\\
+{}&
\widetilde{\Lambda}_{\bm0 \bm q,\downarrow\uparrow}(i\omega_\lambda)
S_{\bm q}^{-}(i\omega_\lambda)
\av{s_{\bm0}^{+}(i\omega_\lambda)}_{\rm FS}
\Bigr],
\label{eq:first_cumulant}
\end{align}
where the spin response associated with the Fermi-surface contribution is defined by
\begin{align}
\av{s_{\bm0}^{\pm}(i\omega_\lambda)}_{\rm FS}
\equiv
\frac{1}{\beta}
\sum_n
\Theta(\varepsilon_n, \omega_\lambda)
\av{s_{\bm0}^{\pm}(i\varepsilon_n,i\omega_\lambda)}_{\rm NM} .
\label{eq:spin_FS}
\end{align}
The expression \eqref{eq:first_cumulant} indicates that the first-order cumulant describes a driving term for the localized spin in the transverse direction.

For the spatially uniform spin response of the NM, \(\bm p=\bm0\), the configurational average of the effective interfacial tunneling correlation is given by $\overline{\mathcal T_{\bm 0,\bm q}}=\mathcal T_1 \delta_{\bm q,\bm 0}$, where $\mathcal T_1$ is the mean of the distribution $\mathcal T_{\bm p,\bm q}$.
In the low-frequency regime, i.e., $\hbar\omega\ll2M$, the frequency dependence of the effective interfacial vertex can be neglected as shown in Eqs.~\eqref{eq:Lambda_retarded1} and \eqref{eq:Lambda_retarded2}.
Performing the analytic continuation $i\omega_\lambda\rightarrow\hbar\omega+i0^+$, the effective magnetic fields are given by
\begin{align}
h^{+}(\omega)
&= \frac{2i}{\hbar\gamma_{\rm g}} \mathcal J_1 \av{s_{\bm0}^{+}(\omega)}_{\rm FS}, 
\label{eq:effective_field_retarded} \\
h^{-}(\omega)
&= - \frac{2i}{\hbar\gamma_{\rm g}} \mathcal J_1^*
\av{s_{\bm0}^{-}(\omega)}_{\rm FS}, 
\label{eq:effective_field_retarded2}
\end{align}
where we define $\mathcal J_1 = \mathcal T_1 \Lambda_{\downarrow}$.
The transverse effective field is written as $\bm h_{\rm eff} = \bm h_{\rm FL} +  \bm h_{\rm DL}$, where
\begin{align}
\bm h_{\rm FL}(\omega) &\equiv -\frac{2}{\hbar\gamma_{\rm g}} (\Im \mathcal J_1) \av{\bm s^{\perp}_{\bm 0}(\omega)}_{\rm FS} + \bm h_{\rm Oe}(\omega),
\\
\bm h_{\rm DL}(\omega) &\equiv \frac{2}{\hbar\gamma_{\rm g}}(\Re \mathcal J_1) \bm e_X \times \av{\bm s^{\perp}_{\bm 0}(\omega)}_{\rm FS}
\label{eq:effective_field_vector}
\end{align}
where $\bm s^{\perp}=s^{Y}\bm e_Y+s^{Z}\bm e_Z$ denotes the spin component transverse to the $X$-direction.
Here, $\bm h_{\rm FL}(\omega)$ and $\bm h_{\rm DL}(\omega)$ represent the effective fields associated with the field-like and damping-like torques, respectively.
The terms ``field-like'' and ``damping-like'' refer to the directions of the corresponding torques and do not imply a temporal phase difference between the two driving fields.
We note that the Oersted field $\bm h_{\rm Oe}(\omega)$ generated by the applied current is incorporated into the field-like driving field.

Together with $h^{\pm}=h^{Y}\pm ih^{Z}$, $S^{\pm}=S^{Y}\pm iS^{Z}$, and $s^{\pm}=s^{Y}\pm i s^{Z}$,  straightforward calculations give
\begin{align}
    h_Y(\omega) &= -\frac{2}{\hbar\gamma_{\rm g}}\left[(\Im \mathcal J_1) \av{s^{Y}_{\bm 0}(\omega)}_{\rm FS} + (\Re \mathcal J_1) \av{s^{Z}_{\bm 0}(\omega)}_{\rm FS}\right] ,
\label{eq:FL_field} 
\\
h_Z(\omega) &= \frac{2}{\hbar\gamma_{\rm g}}\left[(\Re \mathcal J_1) \av{s^{Y}_{\bm 0}(\omega)}_{\rm FS} -(\Im \mathcal J_1)  \av{s^{Z}_{\bm 0}(\omega)}_{\rm FS}\right].
\label{eq:DL_field}
\end{align}
These results show that $\Re \mathcal J_1$ and $-\Im \mathcal J_1$ determine the effective-field components perpendicular and parallel to the transverse spin polarization, respectively. 
For an in-plane spin polarization, i.e., $\av{s^Z_{\bm 0}}=0$, $h_Y$ and $h_Z$ correspond to the field-like and damping-like driving fields, respectively, so that $h_Y=h_{\rm FL}$ and $h_Z=h_{\rm DL}$.

\subsection{FMR frequency shift and damping modulation}

We next consider the second cumulant in Eq.~\eqref{eq:magnon_cumulant}. 
The connected transverse spin correlation of the NM generates a bilinear contribution in the localized-spin operators and thereby renormalizes the magnon propagator. 
We focus on the normal contribution proportional to $S^{+}S^{-}$ and express the second cumulant as
\begin{align}
-\frac{1}{2}
\left\langle
\mathcal S_{\rm int}^{2}
\right\rangle_{\text{NM},c}
=
\beta \sum_{\lambda}
S_{\bm0}^{+}(i\omega_\lambda)
\Sigma_{\bm0}(i\omega_\lambda)
S_{\bm0}^{-}(-i\omega_\lambda),
\label{eq:second_cumulant}
\end{align}
where $\Sigma_{\bm0}(i\omega_\lambda)$ denotes the magnon self-energy generated by the second-order cumulant (see Fig.~\ref{fig:diagram3}(b)).
The explicit form of the self-energy is given by 
\begin{align}
\Sigma_{\bm0}(i\omega_\lambda)
= V_{\rm N} \sum_{\bm p}
\widetilde{\Lambda}_{\bm p\bm0,\downarrow\uparrow}(i\omega_\lambda)
\widetilde{\Lambda}_{-\bm p\bm0,\downarrow\uparrow}(i\omega_\lambda)
\chi_{\bm p}^{\rm FS}(i\omega_\lambda),
\label{eq:magnon_self_energy_matsubara}
\end{align}
where $V_{\rm N}$ is the volume of the NM, and $\chi_{\bm p}^{\rm FS}(i\omega_\lambda)$ is the Fermi-surface contribution to the transverse spin susceptibility, which is defined for $\omega_\lambda>0$ as
\begin{multline}
\chi_{\bm p}^{\rm FS}(i\omega_\lambda)
\equiv
-\frac{1}{\beta V_{\rm N}}
\sum_n
\Theta(\varepsilon_n, \omega_\lambda)
\\
\times
\left\langle
s_{\bm p}^{+}(i\varepsilon_n,i\omega_\lambda)
s_{-\bm p}^{-}(i\varepsilon_n+i\omega_\lambda,-i\omega_\lambda)
\right\rangle_{{\rm NM},c}.
\label{eq:chi_N_matsubara}
\end{multline}
A detailed derivation of Eqs.~\eqref{eq:second_cumulant}--\eqref{eq:chi_N_matsubara} is given in Appendix~\ref{app:magnon_self_energy}.

Considering the low-frequency regime  $\hbar\omega\ll2M$ and performing the analytic continuation $i\omega_\lambda\rightarrow\hbar\omega+i0^{+}$, the retarded self-energy becomes
\begin{align}
\Sigma_{\bm0}^{R}(\omega)
=- N_{\rm FM} V_{\rm N} \Lambda_{\downarrow}^2
\sum_{\bm p}
|\mathcal T_{\bm p,\bm0}|^2
\chi_{\bm p}^{R,{\rm FS}}(\omega),
\label{eq:magnon_self_energy}
\end{align}
where we have used $\mathcal T_{-\bm p,\bm0}=\mathcal T_{\bm p,\bm0}^{*}$.
Thus, in the low-frequency regime, the frequency dependence of the magnon self-energy is governed by the dynamical transverse spin susceptibility of the adjacent electron system.

The dressed magnon Green function for uniform mode is obtained from the Dyson equation,
\begin{align}
G^{R}(\omega)
=
\frac{1}
{\left[G_0^{R}(\omega)\right]^{-1}
-\Sigma_{\bm0}^{R}(\omega)},
\label{eq:magnon_dyson}
\end{align}
where $G_0^{R}(\omega)$ is the unperturbed magnon Green function, given by
\begin{align}
    G_0^{R}(\omega) = \frac{2S/\hbar}{\omega -\omega_{\bm 0}+i\alpha_{\rm G} \omega}.
\end{align}
Here, we have introduced the intrinsic Gilbert damping constant $\alpha_{\rm G}$ for the isolated FM as a phenomenological parameter.
The dressed magnon Green function around the resonance can then be written as
\begin{align}
G^{R}(\omega) =
\frac{2S/\hbar}
{\omega-(\omega_{\bm0}+\delta\omega)
+i(\alpha_{\rm G}+\delta\alpha_{\rm G})\omega}.
\label{eq:dressed_magnon_green}
\end{align}
Then, the resonance-frequency shift and damping modulation are given by
\begin{align}
\frac{\delta\omega}{\omega_{\bm0}} &= - \frac{2SN_{\rm FM}V_{\rm N}}{\hbar\omega_{\bm0}}
\sum_{\bm p}
|\mathcal T_{\bm p,\bm0}|^2
\Re \qty[ \Lambda_{\downarrow}^2 \chi_{\bm p}^{R,{\rm FS}}(\omega_{\bm0}) ], 
\label{eq:res_shift} \\
\delta\alpha_{\rm G} &= \frac{2S N_{\rm FM} V_{\rm N} 
}{\hbar\omega_{\bm0}}
\sum_{\bm p} |\mathcal T_{\bm p,\bm0}|^2 \Im \qty[ \Lambda_{\downarrow}^2 \chi_{\bm p}^{R,{\rm FS}}(\omega_{\bm0}) ].
\label{eq:damp_shift}
\end{align}
Here, we have assumed a sufficiently sharp resonance ($\alpha_{\rm G}+\delta \alpha_{\rm G} \ll 1$), for which the self-energy can be evaluated at the bare resonance frequency $\omega_{\bm0}$. 

These expressions are essentially the same as those for ferromagnetic-insulator/NM junctions~\cite{Ominato2025}, being governed by the transverse spin susceptibility of the NM. 
For $\Gamma_s \ll M$, its real and imaginary parts determine the resonance-frequency shift and damping modulation, respectively, with the NM response restricted to its Fermi-surface contribution. 
A notable feature of the metallic-FM case is the sign reversal arising from the $\pi/2$ phase shift of the effective coupling mediated by the FM conduction electrons.
Although this contribution reduces the Gilbert damping, the lower-order contribution gives a positive damping enhancement and dominates in the weak-tunneling regime, as shown in Appendix~\ref{appsec:spin_pump}. 
Since this lower-order term depends only on the Fermi-level density of states, the present contribution is the leading term that reflects the dynamical spin structure of the NM.

\section{Application to three-dimensional normal metals}
\label{sec:3dNM}

As a simple example, we consider the standard setup of the ST-FMR experiment using a three-dimensional NM with strong spin-orbit interaction, such as platinum.
We assume that the NM occupies the region $-d/2<z<d/2$ and that a spatially uniform ac electric field is applied along the $x$-direction.
The $y$-polarized spin density generated by the spin Hall effect obeys the spin-diffusion equation
\begin{align}
    \left[
    D\partial_z^2 - \partial_t
    -\tau_{\rm sf}^{-1}
    \right]
    \av{s^y(z,t)}
    =0,
    \label{eq:spin-diffusion}
\end{align}
where $D$ is the diffusion constant, and $\tau_{\rm sf}$ is the spin-relaxation time.
The spin current is given by
\begin{align}
    \av{j_{s,z}^y(z,t)} &= j_{\rm SH}(t) - D\partial_z \av{s^y(z,t)}.
\end{align}
Here, $j_{\rm SH}(t)=\theta_{\rm SH}(\hbar/2e) \sigma_{\rm NM} E_x(t)$ is the spin current density driven by the spin Hall effect, $\theta_{\rm SH}$ is the spin Hall angle, $\sigma_{\rm NM}$ is the conductivity of the NM, and $- D\partial_z s^y(z)$ is the diffusive backflow spin current.
The spin Hall effect enters through the boundary condition
$j_{s,z}^y(\pm d/2) = 0$, assuming that the interfacial spin injection is much smaller than $j_{\rm SH}$. 
We assume the condition $\omega \tau_{\rm sf} \ll 1$, which is fulfilled in ST-FMR experiments using platinum.
By solving the spin-diffusion equation under this boundary condition, the steady-state solution is calculated for $\omega\tau_{\rm sf}\ll1$ under ac driving $E_x(t) = E_0 \cos \omega t$ as
\begin{align}
    \av{s^y(z,\omega)} =
    \frac{\lambda_{\rm sf}}{D}
    j_{\rm SH}
    \frac{\sinh\qty(z/\lambda_{\rm sf})}{\cosh \qty(d/2\lambda_{\rm sf})},
    \label{eq:NM_interface_spin}
\end{align}
where $j_{\rm SH}=\theta_{\rm SH}(\hbar/2e) \sigma_{\rm NM} E_0$ is the amplitude of the spin current density, and $\lambda_{\rm sf} = \sqrt{D\tau_{\rm sf}}$ is the spin-diffusion length.
We note that in the low-frequency region $\av{s^y(z,\omega)}$ becomes real and frequency-independent.

The effective magnetic fields are determined by the spin density near the interface $\av{s^y(d/2,\omega)}$.  
The uniform spin
$\av{s^Y_{\bm 0}}_{\rm FS}$ in Eqs.~\eqref{eq:FL_field} and \eqref{eq:DL_field}
is to be replaced by $V_{\rm N}\av{s^Y(d/2)}$, the spin density at the interface
multiplied by the volume of the NM.
Then, the effective fields associated with the field-like and damping-like torques are given by~\cite{funato2020}
\begin{align}
    h_{\rm FL}(\omega) &= -
    \frac{2}{\hbar \gamma_{\rm g}} \frac{(\Im \mathcal J_1) V_{\rm N} j_{\rm SH} \lambda_{\rm sf}}{D}
    \tanh\qty(\frac{d}{2\lambda_{\rm sf}}) \cos \theta,
    \label{eq:NM_interface_spinFL} \\
    h_{\rm DL}(\omega) &= \frac{2}{\hbar \gamma_{\rm g}}
    \frac{(\Re \mathcal J_1) V_{\rm N} j_{\rm SH} \lambda_{\rm sf}}{D}
    \tanh\qty(\frac{d}{2\lambda_{\rm sf}}) \cos \theta.
    \label{eq:NM_interface_spinDL}
\end{align}
Here, we should note that these fields are real and frequency-independent, indicating that they oscillate in phase with the applied electric field.
The resulting dc voltage is therefore given by
\begin{align}
    V_{\rm dc}(\omega)
    &=\frac{R_A I_0\hbar\gamma_g}{8S}
     \sin 2\theta \notag \\
     &\times (h_{\rm FL}
    \Re[ G^R(\omega)]-h_{\rm DL}
    \Im[ G^R(\omega)]),
\end{align}
where we have used $h_Y=h_{\rm FL}$ and $h_Z=h_{\rm DL}$, which hold for an in-plane spin accumulation.
Since both $h_{\rm FL}$ and $h_{\rm DL}$ are proportional to $\cos\theta$,
the dc voltage follows the well-known angular dependence
$V_{\rm dc}\propto \sin 2\theta\cos\theta$~\cite{Liu2011}.
For the weak damping condition, $\Gamma_s\ll M$, the dc voltage is dominated by the damping-like torque and exhibits an approximately Lorentzian line shape around the FMR frequency, $\omega \simeq \omega_{\bm 0}$.
This result is the same as the standard theoretical description of the ST-FMR experiments~\cite{Liu2011}.

\section{Application to 2DEG}\label{sec:2d}

As an illustrative example, we evaluate the dc voltage measured in the ST-FMR experiment for a semiconductor/FM heterostructure, which hosts a 2DEG with Rashba SOC~\cite{yama2021,yama2023,Ominato2025}.
In Secs.~\ref{sec:2DEGmodel} and \ref{sec:FMRmoduation}, we briefly summarize the results for the spin susceptibility of  the 2DEG~\cite{yama2023}. 

\subsection{Model}
\label{sec:2DEGmodel}

The Hamiltonian for a 2DEG with Rashba SOC is given by
\begin{align}
\hat{\mathcal H}_{\rm NM} = \hat{\mathcal H}_{\text{kin}} + \hat{\mathcal H}_{\text{imp}}.
\end{align}
The first term $\hat{\mathcal H}_{\text{kin}}$ represents the kinetic energy and the Rashba SOC:
\begin{align}
\hat{\mathcal H}_{\text{kin}} = \sum_{\bm k, \sigma ,\sigma'} c^\dagger_{\bm k \sigma} \qty[\xi_{\bm k} \hat{I} + \alpha_R (\bm \sigma \times \bm k)_z ]_{\sigma \sigma'} c_{\bm k \sigma'},
\end{align}
where $c_{\bm k \sigma}$ is an annihilation operator for an electron with wavenumber ${\bm k}=(k_x,k_y)$ and spin $\sigma$, $\hat{I}$ is a $2\times 2$ identity matrix, ${\bm \sigma}=(\sigma_x,\sigma_y,\sigma_z)$ are the Pauli matrices, $\xi_{\bm k}=\hbar^2 k^2/2m - \mu$ is the kinetic energy of the conduction electron measured from the Fermi energy, and $\alpha_R$ is the strength of the Rashba SOC.
The second term $\hat{\mathcal H}_{\text{imp}}$ describes the impurity scattering, given by
\begin{align}
\hat{\mathcal H}_{\text{imp}} = \sum_{\bm k,\bm k',\sigma} V_{\bm k-\bm k'} c^\dagger_{\bm k \sigma} c_{\bm k' \sigma},
\end{align}
where $V_{\bm k}$ is the Fourier transform of the nonmagnetic impurity potential $V_{\text{imp}}(\bm r)= u_i \sum_j \delta(\bm r-\bm r_j)$ with $u_i$ the strength of the potential and $\bm r_j$ the position of the $j$-th impurity.

\begin{figure*}[tb]
    \centering
    \includegraphics[width=160mm]{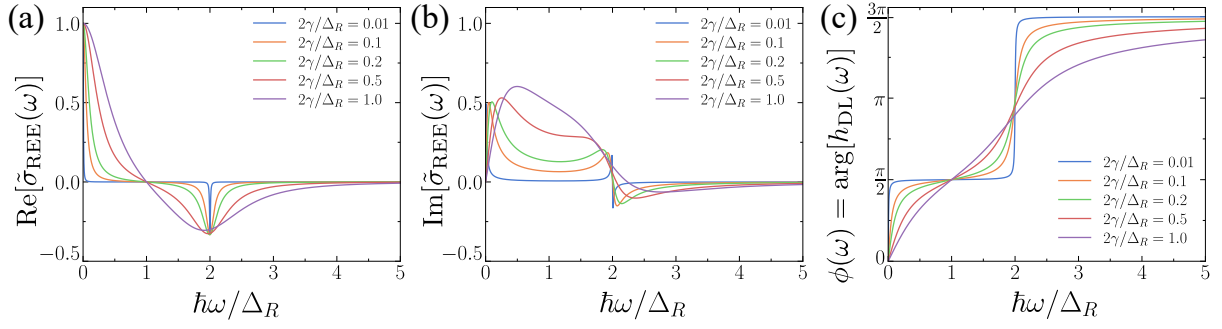}
    \caption{Frequency dependence of the real (a) and imaginary (b) parts of the effective magnetic field associated with the damping-like torque, and its phase \(\phi(\omega)=\arg \qty[h_{\rm DL}(\omega)]\) (c), for different values of the normalized damping rate $2\gamma/\Delta_R$, as indicated by the line colors.
    The field is normalized as $\tilde\sigma_{\rm REE}(\omega)=h_{\rm DL}(\omega)/[h_{{\rm DL},0}\cos\theta]$, so that $\tilde\sigma_{\rm REE}(0)=1$.
    }
    \label{fig:hDL_plot}
\end{figure*}

\subsection{Resonance-frequency shift and damping modulation}
\label{sec:FMRmoduation}

Next, we consider the ST-FMR experiment of the 2DEG.
We assume a clean FM/2DEG interface for which the in-plane component of the electron momentum is conserved and a metallic FM with $\Gamma_s \ll M$.
In this case, we can set the interfacial tunneling correlation as $|\mathcal T_{\bm p,\bm0}|^2 = \mathcal T_1^2 \delta_{{\bm p},{\bm 0}}$.
Using Eqs.~\eqref{eq:res_shift} and \eqref{eq:damp_shift}, the resonance-frequency shift and the Gilbert damping modulation are given as
\begin{align}
\frac{\delta \omega}{\omega_{\bm 0}} &= -A_0 \Re \tilde \chi_{\bm 0}^{R,{\rm FS}}(\omega_{\bm0}), \label{ReTildeChi}\\
\delta \alpha_{\rm G} &= A_0 \Im \tilde \chi_{\bm 0}^{R,{\rm FS}}(\omega_{\bm0}),\label{ImTildeChi}
\end{align}
where $A_0 = 2 \pi SN_{\rm FM} \mathcal J_1^2 \mathcal A D(\varepsilon_{\rm F})/\Delta_R$ is a dimensionless coupling strength, 
$\mathcal A$ is the junction area, $\Delta_{\rm R} = \alpha_R k_{\rm F}$ is the strength of the Rashba SOC, and $\tilde \chi_{\bm 0}^{R,{\rm FS}}(\omega)$ is a normalized spin susceptibility defined by
\begin{align}
    \tilde \chi_{\bm 0}^{R,{\rm FS}}(\omega) = \frac{\Delta_R}{\pi D(\varepsilon_{\rm F}) \hbar \omega} \chi_{\bm 0}^{R,{\rm FS}}(\omega).
\end{align}
Using the ladder approximation, this normalized spin susceptibility is calculated as
\begin{align}
\tilde \chi_{\bm 0}^{R,{\rm FS}}(\omega) &= \frac{1}{8\pi} \sum_{\nu \nu'} 
\qty( \frac{1}{1-\tilde \Gamma_{\parallel}^R(\omega)} 
+ \frac{ 1-\nu\nu' }{1-\tilde \Gamma_{\perp}^R(\omega)})
 \nonumber \\
&\hspace{5mm} \times  \frac{\Delta_{\rm R}}{\hbar \omega - (\nu-\nu')\Delta_{\rm R} + 2i\gamma},
\label{eq:chi2DEG} \\
\tilde \Gamma^R_\parallel(\omega) &=
\frac{i\gamma}{2} \sum_{\nu\nu'} \frac{1}{\hbar \omega - (\nu - \nu') \Delta_{\rm R} + 2i\gamma}, \label{eq:GammaPara} \\
\tilde \Gamma^R_\perp(\omega) &=  i\gamma \sum_{\nu}  \frac{1}{\hbar \omega - 2\nu \Delta_{\rm R} +2i\gamma} ,
\label{eq:GammaPerp}
\end{align}
where $\gamma=\pi n_iu_i^2 D(\varepsilon_{\rm F})$ is the energy broadening due to the impurity scattering.
Detailed derivation is given in Appendix~\ref{appsec:cal_rashba}.
Combining these results with Eqs.~\eqref{ReTildeChi} and \eqref{ImTildeChi}, we can evaluate the FMR frequency shift and damping modulation (for a representative example, see Appendix~\ref{appsec:FMR}).

\subsection{Rashba--Edelstein conductivity}

The Fermi-surface contribution to the spin response is written as
\begin{align}
    \av{\hat s^y(\omega)}_{\rm FS}/\mathcal A
    =
    \sigma_{\rm REE}^{\rm FS}(\omega) E_x ,
\end{align}
which defines the Fermi-surface contribution
$\sigma_{\rm REE}^{\rm FS}(\omega)$ to the Rashba--Edelstein conductivity.
By the linear response theory, the Rashba--Edelstein conductivity is calculated as
\begin{align}
    \sigma_{\rm REE}^{\rm FS}(\omega)
        &= \frac{\sigma_{\rm REE}(0)}{1-\tilde \Gamma_{\parallel}^R(\omega)} \frac{2i\gamma}{\hbar \omega + 2i\gamma}
        \frac{2\Delta_R^2}{ (-i\hbar \omega + 2\gamma)^2 + 4\Delta_R^2}.    
        \label{eq:sigmaREEformula}
\end{align}
For a detailed calculation, see Appendix~\ref{app:REEconductivity}.
In the following, we use a normalized Rashba--Edelstein conductivity $\tilde{\sigma}_{\rm REE}(\omega) = \sigma_{\rm REE}^{\rm FS}(\omega)/\sigma_{\rm REE}(0)$, where $\sigma_{\rm REE}(0) = -e D(\varepsilon_{\rm F}) \alpha_R/2\gamma$ is the conductivity in the dc limit~\cite{edelstein1990}.

\subsection{Effective magnetic field}

Substituting the above result into Eqs.~\eqref{eq:FL_field} and \eqref{eq:DL_field}, we obtain the effective magnetic field as
\begin{align}
    h_{\rm FL}(\omega) &= 
    h_{{\rm FL},0} \cos \theta \, \tilde \sigma_{\rm REE}(\omega),
    \label{eq:rashba_fl}
    \\
    h_{\rm DL}(\omega) &= h_{{\rm DL},0} \cos \theta \, \tilde \sigma_{\rm REE}(\omega),
    \label{eq:rashba_dl}
\end{align} 
where $h_{{\rm DL},0} = (2/\hbar\gamma_{\rm g})\mathcal A E_x \sigma_{\rm REE}(0) \Re \mathcal J_1$
and $h_{{\rm FL},0} = -(2/\hbar\gamma_{\rm g})\mathcal A E_x \sigma_{\rm REE}(0) \Im \mathcal J_1$ represent the effective fields in the dc limit for $\theta=0$.
Under the weak-damping condition, \(\Gamma_s\ll M\), the field-like effective field is much smaller than the damping-like effective field, i.e., $|h_{{\rm FL},0}| \ll |h_{{\rm DL},0}|$. 
We therefore focus on \(h_{\rm DL}(\omega)\) in the following.

Figure~\ref{fig:hDL_plot}(a) and (b) present the frequency dependence of the real and imaginary parts of $h_{\rm DL}(\omega)$, respectively, and Fig.~\ref{fig:hDL_plot}(c) shows the corresponding phase $\phi(\omega) = \arg[h_{\rm DL}(\omega)]$.
In the dc limit, $\hbar\omega=0$, the normalized field is real and equal to unity, $\tilde\sigma_{\rm REE}(0)=1$, so that its phase is zero.
As the frequency increases, the real part decreases and the imaginary part gradually develops, resulting in a finite phase shift of the effective field.
These changes become sharper as the damping is reduced.
At $\hbar\omega=\Delta_R$, the real part vanishes while the imaginary part remains finite.
Hence, the effective magnetic field is purely imaginary with $\phi(\omega)=\pi/2$.
As seen in Fig.~\ref{fig:hDL_plot}(b), the magnitude of the imaginary part increases as the damping $\gamma$ becomes larger.

As the frequency increases further, the real part changes sign and exhibits a negative peak near $\hbar\omega\simeq2\Delta_R$.
This peak originates from the dynamical spin resonance associated with the Rashba spin splitting.
The imaginary part crosses zero and changes sign at $\hbar\omega=2\sqrt{\Delta_R^2+\gamma^2}\simeq2\Delta_R$.
Consequently, the effective magnetic field becomes purely real and negative with $\phi(\omega)=\pi$, after which the phase further evolves toward $3\pi/2$.
The frequency dependence of the spin accumulation generated by the Rashba--Edelstein effect produces a continuous phase rotation of the effective magnetic field with increasing driving frequency. This phase rotation directly affects the line shape of the frequency dependence of the dc voltage generated by ST-FMR.

\subsection{Dc voltage due to ST-FMR}

\begin{figure}[t]
    \centering
    \includegraphics[width=65mm]{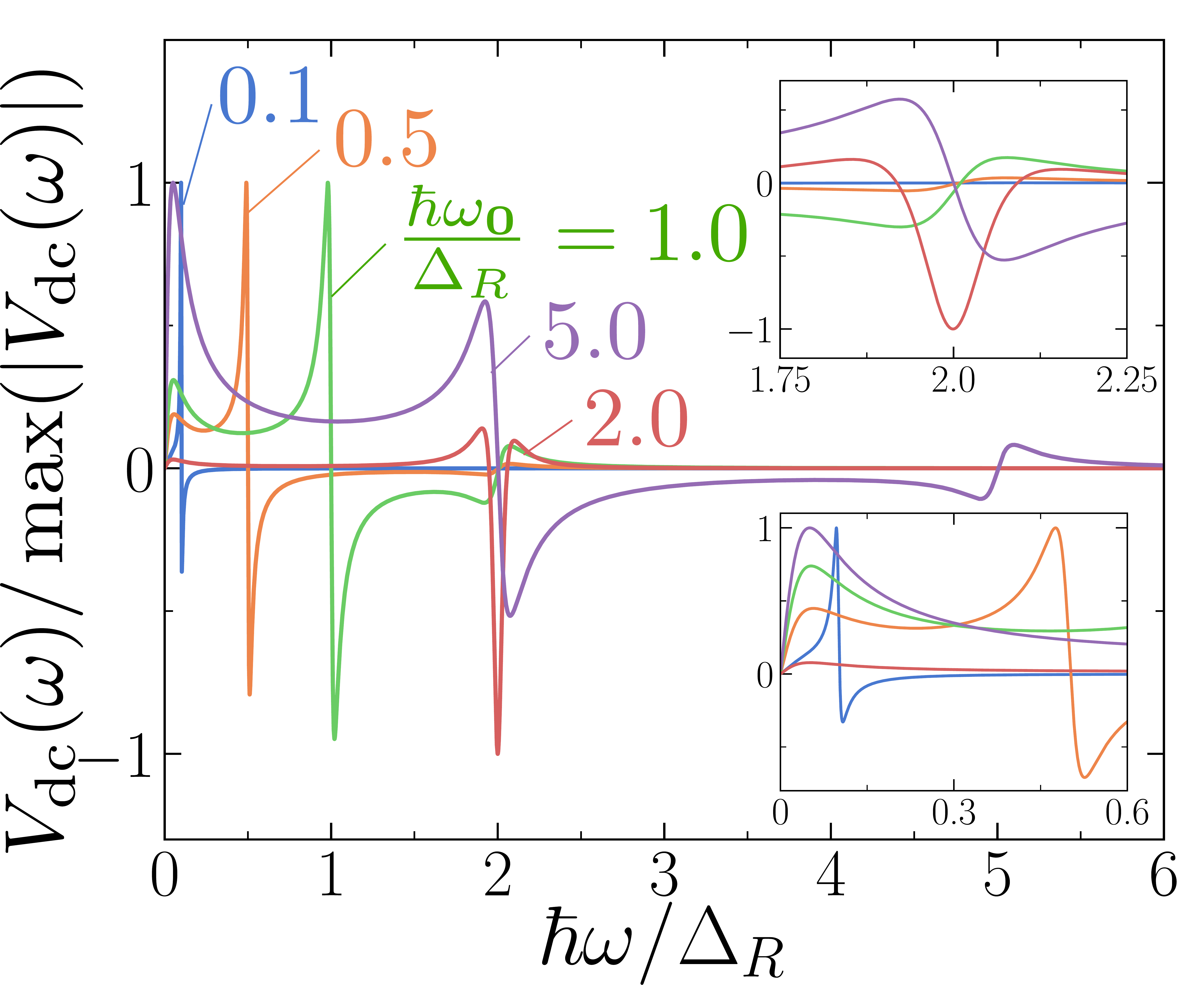}
    \caption{ Frequency dependence of the normalized dc voltage generated by ST-FMR through the damping-like torque for different FMR frequencies.
    Each curve is normalized by its maximum absolute value.
    The insets magnify the low-frequency region and the region around the Rashba resonance $\hbar \omega = 2\Delta_R$.
    We set $\alpha_{\rm eff}=0.05$ and $2\gamma/\Delta_R=0.1$.
    }
    \label{fig:Vdc_plot}
\end{figure}

Neglecting the effective field $h_{\rm FL}$ including the Oersted field and substituting
$h_Z=h_{\rm DL}(\omega)=h_{{\rm DL},0}\cos\theta\,\tilde\sigma_{\rm REE}(\omega)$
into Eq.~\eqref{eq:dc_voltage}, the dc voltage associated with the damping-like torque is expressed as
\begin{align}
     V_{\rm dc} (\omega) &= -V_0 \sin 2\theta \cos\theta
     \Bigl[ \Im[\widetilde G^R (\omega)] \Re[\tilde\sigma_{\rm REE}(\omega)] \notag \\
     & \hspace{20mm} + \Re[\widetilde G^R (\omega)] \Im[\tilde\sigma_{\rm REE}(\omega)]  \Bigr] ,
     \label{eq:Vdc_rashba}
\end{align}
where $V_0 = R_A I_0 \hbar \gamma_g h_{{\rm DL},0}/4 \Delta_R$ represents the amplitude of the dc voltage, $\widetilde{G}^R (\omega) = (\Delta_R/2S) \, G^R (\omega)$ is the dimensionless magnon Green function, and
$\tilde\sigma_{\rm REE}(\omega)=h_{\rm DL}(\omega)/[h_{{\rm DL},0}\cos\theta]$
carries the frequency dependence and the phase of the damping-like field.
The angular dependence $\sin2\theta\cos\theta$ is common to the symmetric and
antisymmetric terms and coincides with that of the conventional ST-FMR signal in
Sec.~\ref{sec:3dNM}.
The phase rotation of the driving field discussed above therefore does not appear in
the angular dependence but manifests itself in the frequency dependence of the line shape.
Provided that the effective magnetic field varies slowly within the FMR linewidth, the first and second terms give the symmetric and antisymmetric line shapes, respectively. 
In contrast to the 3D NM considered in Sec.~\ref{sec:3dNM}, an antisymmetric component develops in the dc-voltage line shape even in the absence of a field-like torque, since the dynamical spin response in the Rashba 2DEG gives rise to a complex driving field.

Figure~\ref{fig:Vdc_plot} presents the driving-frequency dependence of the dc voltage for different FMR frequencies.
Hereafter, we set the total damping constant to be $\alpha_{\rm eff} = \alpha_{\rm G}+\delta\alpha_{\rm G} = 0.05$.
The dc voltage exhibits an FMR feature around \(\omega\simeq\omega_{\bm0}\), with both symmetric and antisymmetric components in the line shape.
The latter appears even for the damping-like torque alone, reflecting the frequency-dependent phase rotation of the effective magnetic field.
The antisymmetric component is particularly pronounced at \(\hbar\omega_{\bm0}=\Delta_R\), where \(\Re [\tilde\sigma_{\rm REE}(\omega)]\) vanishes at resonance, whereas the relative contribution of the symmetric component increases as $\hbar \omega_{\bm 0}$ moves away from $\Delta_R$.
Around \(\hbar\omega_{\bm0}\simeq2\Delta_R\), the symmetric component becomes dominant because the imaginary part of the effective magnetic field is small and the real part dominates at resonance.
Thus, the relative contributions of the real and imaginary parts of the effective magnetic field at resonance depend on the FMR frequency and are reflected in the dc-voltage line shape.

\begin{figure}[t]
    \centering
    \includegraphics[width=65mm]{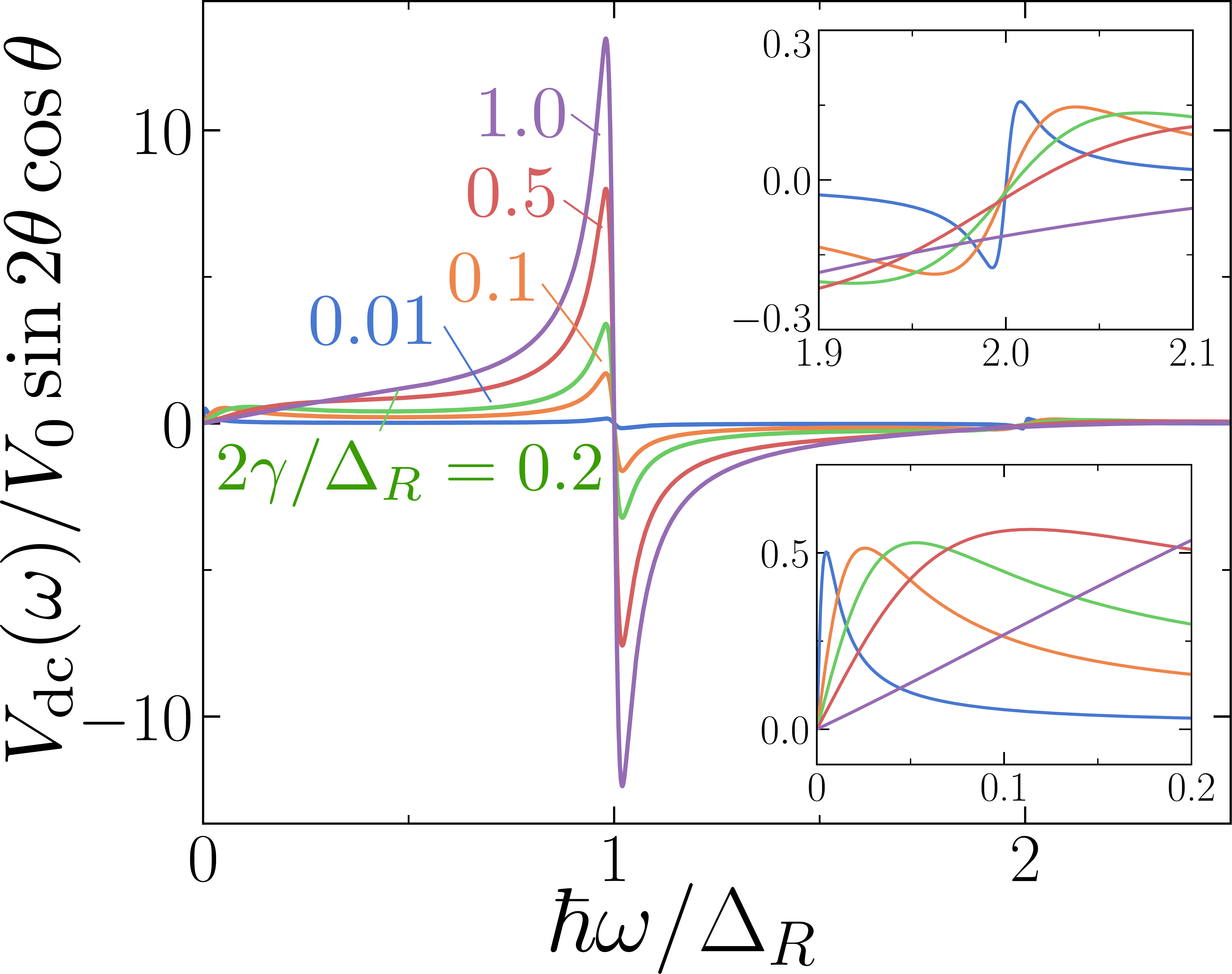}
    \caption{Frequency dependence of the dc voltage generated by ST-FMR through the damping-like torque for different damping rates $2\gamma/\Delta_R$.
    The voltage is normalized by $V_0 \sin 2\theta \cos\theta$ [see Eq.~\eqref{eq:Vdc_rashba}].
    The insets magnify the low-frequency region and the region around the Rashba resonance $\hbar \omega = 2\Delta_R$.
    We set $\alpha_{\rm eff}=0.05$ and $\hbar\omega_{\bm 0}/\Delta_R=1.0$.
    }
    \label{fig:Vdc_gamma}
\end{figure}

As shown in Fig.~\ref{fig:Vdc_gamma}, the dc voltage, in units of
$V_0\sin2\theta\cos\theta$, is plotted as a function of the driving frequency for different damping rates, with the FMR frequency fixed at \(\hbar\omega_{\bm0}=\Delta_R\), where the antisymmetric component is most pronounced.
The dc voltage also shows additional features in the low-frequency region and around the Rashba resonance at \(\hbar\omega\simeq2\Delta_R\), shown in the insets of Fig.~\ref{fig:Vdc_gamma}.
These features mainly originate from the frequency dependence of the effective magnetic field rather than from the magnon resonance.
In the low-frequency region, the dc voltage shows a distinct peak structure.
Here, \(\Im [\tilde\sigma_{\rm REE}(\omega)]\) vanishes as \(\omega\to0\) and rises steeply to a maximum at $\hbar \omega=2\gamma$ [see Fig.~\ref{fig:hDL_plot}(b)], whereas \(\Re \widetilde G^R(\omega)\) remains finite in this frequency range.
The second term of Eq.~\eqref{eq:Vdc_rashba} therefore produces a peak at a low frequency. 
Around \(\hbar \omega \simeq 2\Delta_R\), a zero crossing of the dc voltage occurs when the first and second terms of Eq.~\eqref{eq:Vdc_rashba} cancel each other as the imaginary part of \(\tilde\sigma_{\rm REE}(\omega)\) changes sign.
These features become sharper as the damping rate decreases.
Thus, the dc-voltage spectrum reflects the damping-rate dependence of the dynamical spin response of the NM.

\begin{figure}[tb]
    \centering
    \includegraphics[width=65mm]{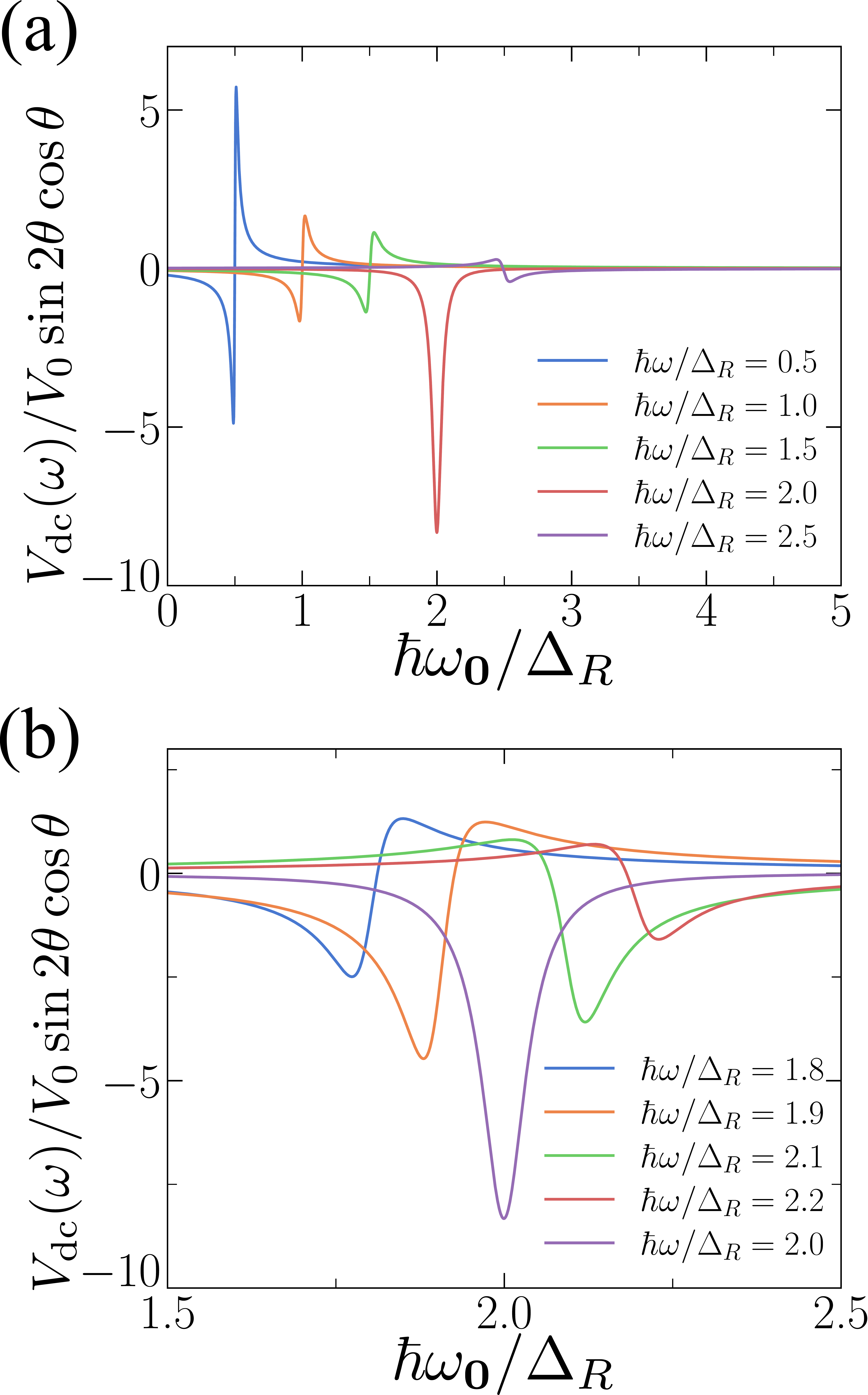}
    \caption{
    (a) Resonance-frequency dependence of the dc voltage generated by ST-FMR through the damping-like torque.
    The voltage is normalized by $V_0 \sin 2\theta \cos\theta$ [see Eq.~\eqref{eq:Vdc_rashba}].
    (b) Enlarged view of the region near $\hbar\omega/\Delta_R=2.0$, with the corresponding values of $\hbar\omega/\Delta_R$ indicated beside each curve.
    We set $\alpha_{\rm eff}=0.05$ and $2\gamma/\Delta_R=0.1$.
    }
    \label{fig:Vdc_res_plot}
\end{figure}

The dependence of the dc voltage on the FMR resonance frequency for different driving frequencies is shown in Fig.~\ref{fig:Vdc_res_plot}.
The dc voltage exhibits a resonance around \(\omega_{\bm0}\simeq\omega\), with the relative contributions of the symmetric and antisymmetric components depending on the driving frequency.
For each driving frequency, the effective magnetic field is fixed, so that the $\omega_{\bm 0}$ dependence mainly arises from the magnon Green function, whereas the relative contributions of its real and imaginary parts are set by the phase of \(h_{\rm DL}\).
Around the Rashba resonance at \(\hbar\omega\simeq2\Delta_R\), the imaginary part of \(\tilde\sigma_{\rm REE}(\omega)\) is small, while its real part is negative. 
The dc voltage therefore exhibits a nearly symmetric FMR resonance with a negative amplitude in units of $V_0\sin2\theta\cos\theta$.
As the driving frequency moves away from \(\hbar\omega=2\Delta_R\), the finite imaginary part of \(\tilde\sigma_{\rm REE}(\omega)\) gives rise to an antisymmetric component in the dc voltage.
Across the Rashba resonance, the sign change of the imaginary part of \(\tilde\sigma_{\rm REE}(\omega)\) reverses the antisymmetric component.
Thus, the phase of \(\tilde\sigma_{\rm REE}(\omega)\) at each driving frequency determines the relative contributions of the symmetric and antisymmetric components to the dc voltage.
The resulting dc-voltage characteristics thus reflect the frequency dependence of the spin accumulation generated by the Rashba--Edelstein effect, suggesting that ST-FMR may provide a means of probing finite-frequency spin generation in the adjacent electron system.
       
\section{Conclusion}\label{sec:conclusion}

We have developed a microscopic formulation of ST-FMR in an NM/FM bilayer system.
Integrating out the FM conduction electrons yields an effective interfacial exchange action between the localized spins in the FM and the conduction electron spins in the NM, with the interfacial tunneling and \(s\)-\(d\) exchange coupling treated perturbatively.
Based on this interaction, we obtained the driving field associated with the current-induced spin-orbit torque, together with the magnon self-energy.
The driving field is governed by the nonequilibrium dynamical spin response of the NM, whereas the magnon self-energy is determined by its dynamical spin susceptibility, with its real and imaginary parts giving the resonance-frequency shift and damping modulation, respectively. Using the resulting driving field and magnon response, we further derived the dc voltage generated by ST-FMR.

We then applied the present formulation to a Rashba 2DEG as a concrete example.
The dynamical spin response of the Rashba 2DEG induces a frequency-dependent phase rotation of the driving field associated with the spin-orbit torque.
This phase rotation gives rise to both symmetric and antisymmetric FMR components in the dc voltage, unlike the predominantly symmetric response in conventional 3D normal metals.
The ST-FMR dc-voltage spectrum thus directly reflects the frequency-dependent spin accumulation generated by the Rashba--Edelstein effect, whose characteristic behavior originates from Rashba spin splitting. More broadly, these results point to ST-FMR as a spectroscopic tool for probing current-induced spin accumulation governed by the band structure and spin texture of adjacent electron systems.

\begin{acknowledgments}
The authors would like to thank Y. Araki and J. Ieda for fruitful discussion.
We are grateful to M. Yama and M. Matsuo for sharing their theoretical expertise.
We also thank H. Nakayama, T. Horaguchi, and T. Kikkawa for helpful discussion from an experimental perspective.
This work was supported by JSPS KAKENHI (Grant Nos. JP24K06951 and JP25KJ0400).
\end{acknowledgments}

\appendix
       
\section{Effective interfacial exchange interaction}
\label{appsec:eff}

In this appendix, we derive the effective interfacial exchange interaction by integrating out the FM conduction electrons. 
Substituting Eqs.~(\ref{eq:fluc_sd}) and (\ref{eq:tunnel}) into Eq.~(\ref{eq:Seff_third_order}) and applying Wick's theorem to the FM conduction-electron operators, we obtain
\begin{align}
\mathcal S_{\rm int} &= - \frac{J_{s\text{-}d}}{2 \sqrt{N_{\rm FM}}} \int ^{ \beta}_0 d\tau_1 d\tau_2 d\tau_3 
\notag \\
&\times 
\sum_{\bm k\bm p\bm l\bm q\sigma}
T^\sigma_{\bm l,\bm k+\bm p}T^{\bar\sigma *}_{\bm l+\bm q,\bm k} S_{\bm q}^{\sigma}(\tau_1)c_{\bm k\bar\sigma}^{\dagger}(\tau _2) c_{\bm k+\bm p\sigma}(\tau_3) \notag \\
& \times  D_{\bm l\sigma}(\tau_1-\tau_3)D_{\bm l+\bm q\bar\sigma}(\tau_2-\tau_1),
\label{apx:int}
\end{align}
where $\sigma=+1$ ($-1$) corresponds to $\uparrow$ ($\downarrow$), $\bar\sigma=-\sigma$, and $S_{\bm q}^{\sigma}$ denotes the raising operator ($\sigma = +1$) or the lowering operator ($\sigma = -1$).
Averaging the second-order tunneling matrix elements over interfacial disorder as $\av{T^\sigma_{\bm l,\bm k+\bm p}T^{\bar\sigma *}_{\bm l+\bm q,\bm k}}_{\rm av}\simeq \mathcal T_{\bm p,\bm q}$ and performing the Fourier transformation, the effective interaction is calculated as
\begin{align}
\mathcal S_{\rm int} 
&= -\frac{J_{s\text{-}d}}{2\sqrt{N_{\rm FM}}}
\sum_{\bm k\bm p\bm l\bm q\sigma}\sum_{n,\lambda}
\mathcal T_{\bm p,\bm q}
S_{\bm q}^{\sigma}(i\omega_\lambda)
c_{\bm k\bar\sigma}^{\dagger}(i\varepsilon_n+i\omega_\lambda)\notag \\
& \hspace{5mm} \times c_{\bm k+\bm p\sigma}(i\varepsilon_n)  D_{\bm l\sigma}(i\varepsilon_n)D_{\bm l+\bm q\bar\sigma}(i\varepsilon_n+i\omega_\lambda).
\label{eq:effective_interaction_matsubara}
\end{align}
Introducing the energy-resolved spin densities of the electrons in the NM as
\begin{align}
s_{\bm p}^{+}(i\varepsilon_n,i\omega_\lambda)
&\equiv \sum_{\bm k}
c_{\bm k\uparrow}^{\dagger}(i\varepsilon_n+i\omega_\lambda)
c_{\bm k+\bm p\downarrow}(i\varepsilon_n),
\\
s_{\bm p}^{-}(i\varepsilon_n,i\omega_\lambda)
& \equiv \sum_{\bm k}
c_{\bm k\downarrow}^{\dagger}(i\varepsilon_n+i\omega_\lambda)
c_{\bm k+\bm p\uparrow}(i\varepsilon_n),
\end{align}
and defining the effective interfacial vertex and the FM conduction-electron kernel as 
\begin{align}
&  \Lambda_{\bm p\bm q,\sigma\bar\sigma}(i\varepsilon_n,i\omega_\lambda) \notag \\
& \equiv \frac{\mathcal T_{\bm p,\bm q} J_{s\text{-}d}}{2\sqrt{N_{\rm FM}}} \sum_{\bm l}
D_{\bm l\sigma}(i\varepsilon_n)D_{\bm l+\bm q\bar\sigma}(i\varepsilon_n+i\omega_\lambda),
\label{eq:lambda_unfactorized}
\end{align}
the effective action can be expressed as
\begin{align}
\mathcal S_{\rm int} 
&= -\sum_{\bm p\bm q}\sum_{n,\lambda}
\Bigl[
\Lambda_{\bm p\bm q}^{\uparrow\downarrow}(i\varepsilon_n,i\omega_\lambda)
S_{\bm q}^{+}(i\omega_\lambda)
s_{\bm p}^{-}(i\varepsilon_n,i\omega_\lambda)
\notag \\
& \hspace{10mm} + \Lambda_{\bm p\bm q}^{\downarrow\uparrow}(i\varepsilon_n,i\omega_\lambda)
S_{\bm q}^{-}(i\omega_\lambda)
s_{\bm p}^{+}(i\varepsilon_n,i\omega_\lambda)
\Bigr].
\end{align}
Note that the effective interfacial coupling depends on both the internal fermionic energy and the transferred bosonic frequency.

We next evaluate the FM conduction-electron kernel within the flat wide-band approximation. 
The momentum sum is replaced by an integration over the energy measured from the Fermi level and the solid angle of the momentum $\bm l$, given in Eq.~\eqref{eq:sum_to_int}.
For $\omega_\lambda>0$, substituting Eq.~\eqref{eq:FM_conduction_green-func} into Eq.~\eqref{eq:lambda_unfactorized}, the momentum sum is evaluated as
\begin{align}
& \frac{1}{N_{\rm FM}}\sum_{\bm l}
D_{\bm l\sigma}(i\varepsilon_n)
D_{\bm l+\bm q,\bar\sigma}(i\varepsilon_n+i\omega_\lambda) \notag \\
& =
\int\frac{d\Omega_{\bm l}}{4\pi}
\frac{2\pi i\tilde{\nu}_{\rm F} \Theta(\varepsilon_n, \omega_\lambda)}{
i\omega_\lambda-2\sigma M-\hbar v_{\rm F}\hat{\bm l}\cdot \bm q - \varepsilon_q+i\Gamma_s
}.
\label{apx:PF_integral}
\end{align}
The kernel is therefore finite only for $-\omega_\lambda<\varepsilon_n<0$, where the two electron propagators have poles in opposite half-planes. Accordingly, the factorized kernel defined in Eq.~\eqref{eq:lambda_factorization} is calculated as
\begin{align}
\widetilde{\Lambda}_{\bm p\bm q,\sigma\bar\sigma}
(i\omega_\lambda) &= \frac{\pi i\tilde{\nu}_{\rm F} \mathcal T_{\bm p,\bm q} \sqrt{N_{\rm FM}} J_{s\text{-}d}}{2\hbar v_{\rm F}q} \notag \\
& \times \ln \left[ \frac{i\omega_\lambda-2\sigma M-\varepsilon_q+i\Gamma_s+\hbar v_{\rm F}q}{i\omega_\lambda-2\sigma M-\varepsilon_q+i\Gamma_s-\hbar v_{\rm F}q } \right].
\label{appeq:lambda_finite_q}
\end{align}
Performing analytic continuation $i\omega_\lambda\rightarrow\hbar\omega+i0^+$ 
and taking the uniform-mode ($q \rightarrow 0$), which is relevant to the spatially uniform FMR excitation typically realized in ST-FMR experiments, we obtain
\begin{align}
\widetilde{\Lambda}_{\bm p\bm0,\sigma\bar\sigma}^{R}(\omega)
= \frac{i\pi \tilde{\nu}_{\rm F} \mathcal T_{\bm p,\bm0} \sqrt{N_{\rm FM}}J_{s\text{-}d}}
{\hbar\omega-2\sigma M+i\Gamma_s}.
\label{appeq:lambda_retarded}
\end{align}

The effective interfacial vertex $\widetilde{\Lambda}_{\bm p\bm0,\sigma\bar\sigma}^{R}(\omega)$ is a frequency-dependent complex number, whose argument can be understood from the equation of motion as follows.
The time evolution of the transverse spin of the conduction electrons in the FM is described by
\begin{align}
\frac{ds_F^+(t)}{dt}
&= \frac{i}{\hbar}[\hat{\mathcal H}_s,s_F^+(t)] - \frac{\Gamma_s}{\hbar} s_F^+(t) + I_T(t) \notag \\
& =\frac{2M - i\Gamma_s}{i\hbar}s_F^+(t) + I_T(t), 
\end{align}
where $s_F^+ = \sum_{\bm l} a_{\bm l \uparrow}^\dagger a_{\bm l \downarrow}$, $I_T(t)$ is the spin-injection rate from the NM, and $\Gamma_s$ is the transverse spin broadening.
Fourier transforming this equation yields
\begin{align}
s_F^+(\omega)=\frac{i\hbar}{\hbar\omega-2M+i\Gamma_s}I_T(\omega).
\label{apx:EOM_solution}
\end{align}
In the lowest-order approximation for the interfacial electron hopping, the injection rate $I_T(\omega)$ becomes a real number.
Therefore, the argument of $s_F^+(\omega)$ has the same frequency dependence as that of $\widetilde{\Lambda}_{\bm p\bm0,\sigma\bar\sigma}^{R}(\omega)$ given in Eq.~\eqref{appeq:lambda_retarded}. 
In the low-frequency and weak-relaxation limit, $\hbar\omega, \Gamma_s \ll 2M$, the transverse spin becomes $s_F^+(\omega)\simeq - (i\hbar/2M) I_T(\omega)$.
This predominantly imaginary response represents the $\pi/2$ phase shift associated with the precession of the transverse spin around the exchange field.

\section{Calculation of magnon self-energy}
\label{app:magnon_self_energy}

In this appendix, we derive the magnon self-energy from the second-order cumulant of the effective interfacial interaction and obtain the resulting resonance-frequency shift and damping modulation. 
The connected part of the second-order cumulant is written as
\begin{align}
-\frac{1}{2}\left\langle
\mathcal S_{\rm int}^{2}
\right\rangle_{{\rm NM},c} 
& = -\frac{\beta^2}{2}\sum_{\lambda} \sum_{\eta}
S_{\bm0}^{\eta}(i\omega_\lambda)
S_{\bm0}^{\bar{\eta}}(-i\omega_{\lambda}) \notag \\
& \hspace{5mm} \times \left\langle
\mathcal B^{\bar{\eta}}(-i\omega_\lambda)
\mathcal B^{\eta}(i\omega_{\lambda})
\right\rangle_{{\rm NM},c},
\label{eq:second_cumulant_B}
\end{align}
where $\eta = \pm$ assigns the raising and lowering spin operators, $\bar{\eta} = \mp$ indicates the opposite operation, and $B^{\eta}(i\omega_\lambda)$ is defined by
\begin{align}
\mathcal B^{\eta}(i\omega_\lambda)
&\equiv
\frac{1}{\beta}
\sum_{\bm p,n}
\Lambda_{\bm p\bm0,\bar{\eta}\eta}
(i\varepsilon_n,i\omega_\lambda)
s_{\bm p}^{\eta}(i\varepsilon_n,i\omega_\lambda),
\label{eq:B_definition}
\end{align}
with $\eta=+,(-)$ identified with $\uparrow,(\downarrow)$ for $\Lambda_{\bm p\bm0,\bar{\eta} \eta} (i\varepsilon_n,i\omega_\lambda)$.
We note that anomalous quadratic terms proportional to $S^{+}S^{+}$ or $S^{-}S^{-}$ have been neglected, as they do not contribute to the magnon self-energy in the leading order of the interfacial coupling.
The average for the connected diagram is calculated as
\begin{align}
&\left\langle\mathcal B^{+}(i\omega_\lambda)\mathcal B^{-}(-i\omega_\lambda)\right\rangle_{{\rm NM},c} \notag \\
&= \frac{1}{\beta^{2}}\sum_{\bm p,n}
\Lambda_{\bm p\bm0,\downarrow\uparrow}(i\varepsilon_n,i\omega_\lambda)
\Lambda_{-\bm p\bm0,\uparrow\downarrow}(i\varepsilon_n+i\omega_\lambda,-i\omega_\lambda)
\notag \\
& \hspace{5mm} \times \left\langle
s_{\bm p}^{+}(i\varepsilon_n,i\omega_\lambda)
s_{-\bm p}^{-}(i\varepsilon_n+i\omega_\lambda,-i\omega_\lambda)
\right\rangle_{{\rm NM},c}.
\label{eq:BplusBminus_FS}
\end{align}
Using Eq.~\eqref{eq:lambda_unfactorized} together with the flat wide-band approximation, we obtain
\begin{align}
& \Lambda_{\bm p\bm0,\downarrow\uparrow}(i\varepsilon_n,i\omega_\lambda)
=\Theta(\varepsilon_n, \omega_\lambda)
\widetilde{\Lambda}_{\bm p\bm0,\downarrow\uparrow}(i\omega_\lambda),
\\
& \Lambda_{-\bm p\bm0,\uparrow\downarrow}(i\varepsilon_n+i\omega_\lambda,-i\omega_\lambda) \notag \\
& \hspace{24mm} =\Theta(\varepsilon_n, \omega_\lambda) \widetilde{\Lambda}_{-\bm p\bm0,\downarrow\uparrow}(i\omega_\lambda).
\label{eq:Lambda_second_cumulant}
\end{align}
Combining Eqs.~\eqref{eq:BplusBminus_FS}-\eqref{eq:Lambda_second_cumulant}, the second-order cumulant is calculated as 
\begin{align}
-\frac{1}{2}
\left\langle
\mathcal S_{\rm int}^{2}
\right\rangle_{\text{NM},c}
=
\beta
\sum_{\lambda}
S_{\bm0}^{+}(i\omega_\lambda)
\Sigma_{\bm0}(i\omega_\lambda)
S_{\bm0}^{-}(-i\omega_\lambda),
\label{eq:second_cumulant_self_energy}
\end{align}
where the magnon self-energy is given by
\begin{align}
\Sigma_{\bm0}(i\omega_\lambda)
=
V_{\rm N} \sum_{\bm p}
\widetilde{\Lambda}_{\bm p\bm0,\downarrow\uparrow}(i\omega_\lambda)
\widetilde{\Lambda}_{-\bm p\bm0,\downarrow\uparrow}(i\omega_\lambda)
\chi_{\bm p}^{\rm FS}(i\omega_\lambda).
\label{eq:magnon_self_energy_matsubara2}
\end{align}
Here, $\chi_{\bm p}^{\rm FS}(i\omega_\lambda)$ is the Fermi-surface contribution to the transverse spin susceptibility of the NM, given in Eq.~\eqref{eq:chi_N_matsubara}.

\section{Leading-order contribution to magnon self-energy}
\label{appsec:spin_pump}

\begin{figure}
    \centering
    \includegraphics[width=0.9\linewidth]{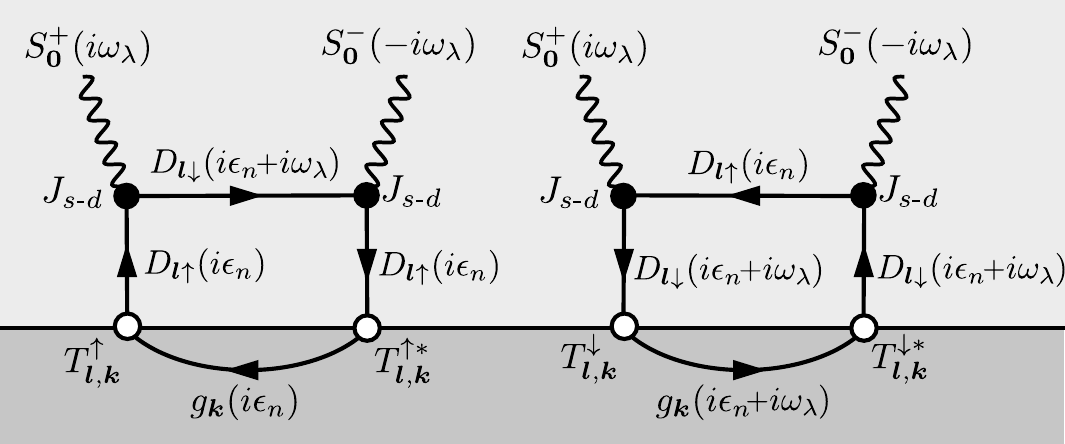}
    \caption{Diagrammatic representation of $\av{\mathcal S_{\rm int}^{(2)}}_{\rm NM,c}$ involving two $s$-$d$ exchange interactions and two interfacial electron-tunneling processes.
    The left and right diagrams correspond to the first and second terms in Eq.~\eqref{apx:spin_pumping_self_energy}, respectively.}
    \label{fig:diagram_pump}
\end{figure}

In this appendix, we derive the leading contribution in the tunneling expansion of the magnon self-energy, proportional to $J_{s\text{-}d}^2T^2$, together with the resulting resonance-frequency shift and damping modulation.
The corresponding effective action is given by
\begin{align}
 \mathcal S_{\rm int}^{(2)}
&= -\frac{1}{2! 2!}
\int_0^{\beta}d\tau_1\,d\tau_2\,d\tau_3\, d\tau_4
\notag \\
&\qquad
\times \left\langle
T_\tau
\hat H_{s\text{-}d}(\tau_1)
\hat H_{s\text{-}d}(\tau_2)
\hat H_t(\tau_3)
\hat H_t(\tau_4)
\right\rangle_{s,c}.
\end{align}
Neglecting the terms proportional to $S^{+}S^{+}$ or $S^{-}S^{-}$ and focusing on the uniform localized-spin mode, the effective action reduces to
\begin{align}
\mathcal S_{\rm int}^{(2)}
=&
\frac{J_{s-d}^{\,2}}{4N_{\rm FM}}
\int_0^{\beta}
d\tau_1\,d\tau_2\,d\tau_3\,d\tau_4
\sum_{\bm k,\bm p,\bm l,\sigma}
T_{\bm l,\bm k+\bm p}^{\sigma}
T_{\bm l,\bm k}^{\sigma *}
\notag\\
&\times
S_{\bm 0}^{\sigma}(\tau_1)
S_{\bm 0}^{\bar\sigma}(\tau_2)
c_{\bm k\sigma}^{\dagger}(\tau_3)
c_{\bm k+\bm p,\sigma}(\tau_4)
\notag\\
&\times
D_{\bm l\sigma}(\tau_1-\tau_4)
D_{\bm l\bar\sigma}(\tau_2-\tau_1)
D_{\bm l\sigma}(\tau_3-\tau_2),
\end{align}
with \(S_{\bm 0}^{\sigma}=S_{\bm 0}^{+}\) (\(S_{\bm 0}^{-}\)) for \(\sigma=\uparrow\) (\(\downarrow\)).
Tracing out the NM conduction electrons to first order in $\mathcal S_{\rm int}^{(2)}$ gives
\begin{align}
\av{ \mathcal S_{\rm int}^{(2)}}_{\text{NM},c}
=
\beta \sum_{\lambda}
S_{\bm0}^{+}(i\omega_\lambda)
\Sigma_{\rm sp}(i\omega_\lambda)
S_{\bm0}^{-}(-i\omega_\lambda),
\label{apx:second_order_magnon_action}
\end{align}
where the corresponding diagrams are shown in Fig.~\ref{fig:diagram_pump}. 
The magnon self-energy is given by
\begin{align}
&\Sigma_{\rm sp}(i\omega_\lambda)
=
\frac{J_{s\text{-}d}^{2}}
{4N_{\rm FM}\beta}
\sum_{\bm l,\bm k,n}
\notag \\
& 
\times \Big[
|T^\uparrow_{\bm l,\bm k}|^2
g_{\bm k}(i\varepsilon_n)
D_{\bm l\uparrow}^{2}(i\varepsilon_n)
D_{\bm l\downarrow}(i\varepsilon_n+i\omega_\lambda)
\notag\\
&
+
|T^\downarrow_{\bm l,\bm k}|^2 g_{\bm k}(i\varepsilon_n+i\omega_\lambda)
D_{\bm l\downarrow}^{2}(i\varepsilon_n+i\omega_\lambda)
D_{\bm l\uparrow}(i\varepsilon_n)
\Big].
\label{apx:spin_pumping_self_energy}
\end{align}
Within the flat wide-band approximation \eqref{eq:sum_to_int}, the FM momentum sum is evaluated as
\begin{align}
\frac{1}{N_{\rm FM}}
\sum_{\bm l}
D_{\bm l\uparrow}^{2}(i\varepsilon_n)
D_{\bm l\downarrow}(i\varepsilon_n+i\omega_\lambda)
=
-\frac{2\pi i\widetilde\nu_F
\Theta(\varepsilon_n, \omega_\lambda)}
{(i\omega_\lambda-2M+i\Gamma_s)^2}.
\label{apx:DDD}
\end{align}
For the NM conduction electrons, the same approximation gives $N_{\rm NM}^{-1}\sum_{\bm k}g_{\bm k}(i\epsilon_n) = -i\pi \tilde \nu_N \text{sgn}\,(\epsilon_n)$, where $\tilde \nu_N$ is the density of states per site per spin at the Fermi level for the conduction electrons in the NM and $N_{\rm NM}$ is the number of sites in the NM.
With the disorder-averaged tunneling amplitude
$|T^\uparrow_{\bm l,\bm k}|^2+|T^\downarrow_{\bm l,\bm k}|^2=2|T|^2$,
the retarded self-energy for $\hbar\omega\ll2M$ is obtained as
\begin{align}
\Sigma_{\rm sp}^R(\omega)
=
-i\frac{\hbar \omega }{2S} N_{\rm NM} |T|^2
\frac{\pi \widetilde\nu_F \tilde \nu_N}{S} 
\frac{4M^2}{(2M - i\Gamma_s)^2}.
\label{apx:spin_pumping_self_energy_final}
\end{align}
The resulting resonance-frequency shift and damping modulation are
\begin{align}
\frac{\delta\omega}{\omega_{\bm 0}}
&=
N_{\rm NM} |T|^2
\frac{\pi\widetilde\nu_F\tilde \nu_N}{S}
\frac{16M^3\Gamma_s}
{\left(4M^2+\Gamma_s^2\right)^2},
\\
\delta\alpha_G
&=
N_{\rm NM} |T|^2
\frac{\pi\widetilde\nu_F\tilde \nu_N}{S}
\frac{4M^2\left(4M^2-\Gamma_s^2\right)}
{\left(4M^2+\Gamma_s^2\right)^2},
\end{align}
where the self-energy is evaluated at
$\omega\simeq\omega_{\bm0}$ assuming a sufficiently sharp resonance.
The present contribution gives a positive Gilbert-damping enhancement for metallic FMs satisfying $\Gamma_s<2M$. 
It originates from single-electron tunneling across the NM/FM interface and involves only a single NM electron propagator\cite{Tserkovnyak2002,tatara2017}, in contrast to the contributions in Eqs.~\eqref{eq:res_shift} and \eqref{eq:damp_shift}, which are governed by the NM spin susceptibility. 
It is therefore absent in ferromagnetic-insulator/NM systems. 
Within the flat wide-band approximation, the NM propagator reduces to the Fermi-level DOS, so that this contribution does not reflect the dynamical spin resonance of the NM.

\section{Detailed calculation for the Rashba system}
\label{appsec:cal_rashba}

In this appendix, we show the detailed calculation of the spin susceptibility~\cite{yama2021} and the Edelstein conductivity.
Throughout this paper, the Rashba SOC coupling $\Delta_{\rm R}=\alpha_R k_{\rm F}$ is assumed to be much smaller than the Fermi energy and much larger than the energy broadening due to the impurities $\gamma$.

\subsection{Green's function}

We first apply the Born approximation to treat the impurity scattering described by $\hat{\cal H}_{\rm imp}$.
Assuming a uniformly random distribution of the impurities~\cite{BruusFlensberg2004}, the impurity-averaged Matsubara Green function is given by the $2\times 2$ matrix form~\cite{yama2021}
\begin{align}
g_{\bm k}(i\varepsilon_n) = \sum_{\nu=\pm} \frac{\mathcal P_{\nu}}{i\varepsilon_n - E_{\bm k\nu} + i\gamma \sgn(\varepsilon_n)}, 
\label{eq:Green_g}
\end{align}
where $E_{\bm k\nu}= \xi_{\bm k} + \nu \alpha_R k$ is the eigenenergy of the band $\nu$, $\gamma=\pi n_iu_i^2 D(\varepsilon_{\rm F})$ is the energy broadening due to the impurity scattering, $n_i$ is the impurity concentration,  $D(\varepsilon_{\rm F}) = m/2\pi \hbar^2$ is the density of states per spin at the Fermi level in the absence of the SOC, and $\mathcal P_\nu = [\hat{I} - \nu(\hat{\bm k}\times\bm\sigma)_z]/2$ is the band-projection operator.

\subsection{Ladder vertex correction}

\begin{figure}[t]
    \centering
    \includegraphics[width=55mm]{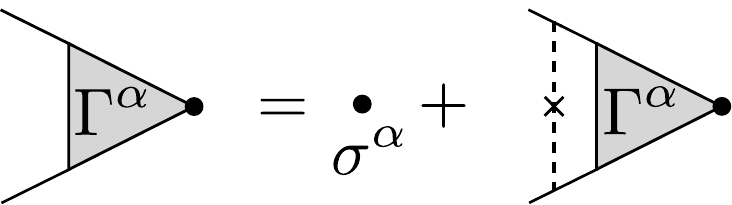}
    \caption{Feynman diagram of the Bethe--Salpeter equation for the ladder approximation.
    The straight and dashed lines represent the electron propagator and  the impurity scattering, respectively.}
    \label{fig:Feynman}
\end{figure}

It is useful to introduce the Pauli matrices in the magnetization-fixed coordinates (see also Fig.~\ref{fig:fig2}(b)), which are defined as
\begin{align}
\left( \begin{array}{c} \sigma^X \\ \sigma^Y \\ \sigma^Z \end{array} \right) =
\left( \begin{array}{ccc} \cos \theta & \sin \theta & 0 \\ -\sin \theta & \cos \theta & 0 \\
0 & 0 & 1 \end{array} \right) 
\left( \begin{array}{c} \sigma^x \\ \sigma^y \\ \sigma^z \end{array} \right) .
\end{align}
For 2DEG with the SOC, the spin susceptibility defined in Eq.~(\ref{eq:chi_N_matsubara}) is calculated as
\begin{align}
\chi_{\bm 0}^{\rm FS} (i\omega_\lambda) &= \frac{1}{4\beta \mathcal A} \sum_{\bm k,n} 
\Theta(\varepsilon_n, \omega_\lambda) \, 
{\rm Tr} \, [ \Gamma^+(i\varepsilon_n+i\omega_\lambda,i\varepsilon_n) \notag \\
& \hspace{15mm} \times g_{\bm k}(i\varepsilon_n+i\omega_\lambda)  \sigma^- g_{\bm k}(i\varepsilon_n) ], \label{eq:chi0FSimag} 
\end{align}
where $\sigma^- = \sigma^Y - i \sigma^Z$, $\Gamma^+ = \Gamma^Y+i\Gamma^Z$, and $\Gamma^\alpha(i\varepsilon_n+i\omega_\lambda,i\varepsilon_n)$ ($\alpha = X, Y, Z$) are $2\times 2$ matrices, which represent the vertex functions due to the impurities.
In this study, we evaluate the vertex functions within the ladder approximation.

In our study, Green's function for conduction electrons in 2DEG has been calculated using the Born approximation (see Eq.~\eqref{eq:Green_g}).
The corresponding vertex corrections are provided by the ladder approximation, whose Bethe-Salpeter equation is given by~\cite{yama2021}
\begin{align}
    &\Gamma^{\alpha}(i\varepsilon_n+i\omega_\lambda, i\varepsilon_n) = \sigma^\alpha \notag \\
    &+ \frac{n_iu_i^2}{\mathcal A} \sum_{\bm k} g_{\bm k}(i\varepsilon_n) \Gamma^\alpha (i\varepsilon_n+i\omega_\lambda, i\varepsilon_n) g_{\bm k} (i\varepsilon_n + i\omega_\lambda).
\end{align}
The Feynman diagram for this Bethe--Salpeter equation is shown in Fig.~\ref{fig:Feynman}.
We expand the ladder vertex with the Pauli matrices as
\begin{align}
    \Gamma^{\alpha}(i\varepsilon_n+i\omega_\lambda, i\varepsilon_n) = \sum_{\beta} \Gamma^\alpha_{\beta}(i\varepsilon_n+i\omega_\lambda, i\varepsilon_n) \sigma^\beta,
\end{align}
where $\Gamma^\alpha_{\beta} = 
{\rm Tr} \, [\Gamma^{\alpha} \sigma^\beta]/2$.
Substituting this expansion into the Bethe--Salpeter equation, the coefficients satisfy 
\begin{align}
    & \Gamma^\alpha_{\beta}
    =\delta_{\alpha\beta} + n_iu_i^2 \sum_{\nu \nu'} D_\nu(\varepsilon_{{\rm F}}) \int dE_{\bm k\nu} \int\frac{d\varphi}{2\pi}  \notag \\
    & \hspace{20mm} \times \frac{\sum_{\gamma} \Gamma^\alpha_{\gamma}\Tr[\sigma^\beta \mathcal P_{\nu'} \sigma^\gamma \mathcal P_{\nu}]}{2\mathcal D_{\bm k\nu}(i\varepsilon_n+i\omega_\lambda) \mathcal D_{\bm k\nu'}(i\varepsilon_n) } ,
\end{align}
where $\delta_{\alpha\beta}$ is the Kronecker delta, $D_\nu(\varepsilon_{{\rm F}})\simeq D(\varepsilon_{\rm F})(1-\nu \Delta_R/2\varepsilon_{\rm F})$ is the density of states at the Fermi energy for the band $\nu$, $\mathcal D_{\bm k\nu}(i\varepsilon_n)= i\varepsilon_n - E_{\bm k\nu} + i\gamma \sgn(\varepsilon_n)$ is the denominator of the Green's function of the electrons in the Rashba system (see Eq.~\eqref{eq:Green_g}), and the summation over the wavenumber ${\bm k}$ is replaced with the integral as
\begin{align}
    \frac{1}{\mathcal A}\sum_{\bm k} (\cdots) \rightarrow 
    D_\nu(\varepsilon_{\rm F}) \int dE_{\bm k\nu} \int\frac{d\varphi}{2\pi} (\cdots) .
\end{align}

For a positive bosonic Matsubara frequency, \(\omega_\lambda>0\), the energy integral has a finite contribution only from the retarded-advanced sector, namely \(\varepsilon_n<0\) and \(\varepsilon_n+\omega_\lambda>0\). 
In this condition, the energy integration is evaluated as
\begin{align}
    &\int ^\infty_{-\infty} \frac{dE_{\bm k\nu}}{ \mathcal D_{\bm k\nu}(i\varepsilon_n+i\omega_\lambda) \mathcal D_{\bm k\nu'}(i\varepsilon_n) }\notag \\
        &\qquad = \frac{2\pi i\Theta(\varepsilon_n, \omega_\lambda)}{  i\omega_\lambda - (\nu -\nu') \Delta_R + 2i\gamma  },
\end{align}
and the angular average over the direction of the wave vector is calculated as
\begin{align}
   \int\frac{d\varphi}{2\pi}  \Tr[\sigma^\beta \mathcal P_{\nu'} \sigma^\gamma \mathcal P_{\nu}]
    = \frac{1}{2}\delta_{\beta \gamma} \qty[ 1 - \nu \nu'\delta_{\beta Z} ].
    \label{app:angle_av}
\end{align}
Thus, the ladder equation becomes diagonal in spin space. 
The coefficients of the ladder vertex satisfy 
\begin{align}
    \Gamma^\alpha_{\beta}
    &=\delta_{\alpha \beta} + \Gamma^\alpha_{\beta} \sum_{\nu \nu'} \frac{i\gamma \Theta(\varepsilon_n, \omega_\lambda)/2}{  i\omega_\lambda - (\nu -\nu') \Delta_R + 2i\gamma } \qty[ 1 - \nu \nu'\delta_{\beta Z} ].
\end{align}
Solving this equation, we obtain the ladder vertex corrections as
\begin{align}
    & \Gamma^\alpha (i\varepsilon_n+i\omega_\lambda, i\varepsilon_n)  = 
    \begin{cases}
        \dfrac{\sigma^\alpha}{1-\tilde \Gamma_{\parallel}(i\omega_\lambda)}, & (\alpha = X, Y),
        \\[10pt]
        \dfrac{\sigma^\alpha}{1-\tilde \Gamma_{\perp}(i\omega_\lambda)},  & (\alpha = Z).
    \end{cases} \\
& \tilde \Gamma_\parallel(i\omega_\lambda) = \frac{i\gamma}{2} \sum_{\nu\nu'}
\frac{1}{i\omega_\lambda - (\nu - \nu') \Delta_R + 2i\gamma}, \\
& \tilde \Gamma_\perp(i\omega_\lambda) =  i\gamma \sum_{\nu}  \frac{1}{i\omega_\lambda - 2\nu \Delta_R +2i\gamma} ,
\end{align}
where the functions $\tilde{\Gamma}_{\parallel}$ and $\tilde{\Gamma}_{\perp}$ represent the ladder part of the Feynman diagram for the in-plane ($XY$) and out-of-plane ($Z$) spin components.
The analytic continuation $i\omega_\lambda \rightarrow \hbar \omega + i\delta$ gives the expressions, Eqs.~\eqref{eq:GammaPara} and \eqref{eq:GammaPerp}.

\subsection{Spin susceptibility}

The spin susceptibility is calculated as
\begin{align}
    &\chi_{\bm 0}^{{\rm FS}}(i\omega_\lambda) 
    = \frac{1}{4}\sum_{\nu \nu'} D_{\nu}(\varepsilon_{\rm F}) 
    \Pi_{\nu\nu'}(i\omega_\lambda) \notag \\
    & \times \int \frac{d\varphi}{2\pi} \Bigl[ 
    \frac{ \Tr [\sigma^Y \mathcal P_{\nu} \sigma^- \mathcal P_{\nu'}]}{1-\tilde \Gamma_{\parallel}(i\omega_\lambda)} 
    + \frac{i \Tr [\sigma^Z \mathcal P_{\nu} \sigma^- \mathcal P_{\nu'}]}{1-\tilde \Gamma_{\perp}(i\omega_\lambda)}
    \Bigr],
\end{align}
where $\Pi_{\nu\nu'}(i\omega_\lambda)$ represents the Matsubara summation and the energy integration of the two electron propagators defined by
\begin{align}
    \Pi_{\nu\nu'}(i\omega_\lambda)
    = \frac{1}{\beta}\sum_n \int dE_{\bm k\nu} \frac{\Theta(\varepsilon_n, \omega_\lambda)}{\mathcal D_{\bm k\nu}(i\varepsilon_n + i\omega_\lambda) \mathcal D_{\bm k\nu'}(i\varepsilon_n)}.
    \label{app:Pi-def}
\end{align}
The angular average over the direction of the wave vector is calculated as
\begin{align}
    \int \frac{d\varphi}{2\pi} \, \Tr [\sigma^Y \mathcal P_{\nu} \sigma^- \mathcal P_{\nu'}] &= \frac{1}{2},
    \\
    \int \frac{d\varphi}{2\pi} \, \Tr [i\sigma^Z \mathcal P_{\nu} \sigma^- \mathcal P_{\nu'}] &= \frac{1}{2}(1-\nu \nu').
\end{align}
By performing the analytic continuation $i\omega_\lambda\rightarrow\hbar\omega+i0^+$, the retarded component of the spin susceptibility is given by
\begin{align}
    &\chi^{R,{\rm FS}}_{\bm 0}(\omega) \notag \\
    &=\sum_{\nu \nu'} \frac{D_{\nu}(\varepsilon_{\rm F})}{8}
    \qty(
        \frac{1}{1-\tilde \Gamma_{\parallel}^R(\omega)} 
    + \frac{ 1-\nu\nu' }{1-\tilde \Gamma_{\perp}^R(\omega)}) \Pi_{\nu\nu'}^R(\omega) .
    \label{app:chifspre}
\end{align}

Next, we calculate $\Pi_{\nu\nu'}^R(\omega)$.
By performing the analytic continuation $i\omega_\lambda \to \hbar \omega +i0^+$, it represents the Fermi-surface contribution as
\begin{align}
    \Pi_{\nu \nu'}^R(\omega) &= - \int dE_{\bm k\nu}\int^\infty_{-\infty} \frac{d\varepsilon}{2\pi i} \frac{f(\varepsilon +\hbar \omega) - f(\varepsilon)}{\mathcal D^R_{\bm k\nu}(\varepsilon + \hbar \omega) \mathcal D^{A}_{\bm k\nu'}(\varepsilon)}.
\end{align}
Substituting $\mathcal D^R_{\bm k\nu}(\varepsilon)= \mathcal D^A_{\bm k\nu}(\varepsilon)^*= \varepsilon - E_{\bm k\nu} + i\gamma $, we obtain
\begin{align}
    & \Pi_{\nu \nu'}^R(\omega) = -\int dE_{\bm k\nu}\int^\infty_{-\infty} \frac{d\varepsilon}{2\pi i} (f(\varepsilon +\hbar \omega) - f(\varepsilon)) \notag \\
    & \hspace{15mm} \times \frac{1}{(\varepsilon + \hbar \omega - E_{\bm k \nu} + i\gamma)(\varepsilon - E_{\bm k \nu'} - i\gamma)} 
    \notag \\
    &= -\int dE_{\bm k\nu}\int^\infty_{-\infty} \frac{dE} {2\pi i} (f(-E+E_{\bm k\nu}) - f(E+E_{\bm k\nu'}))  \notag \\
    & \hspace{5mm} \times \frac{1}{E-i\gamma} \frac{1}{E+\hbar \omega - (E_{\bm k\nu} - E_{\bm k\nu'} ) + i\gamma}.
\end{align}
In the second line, we replaced the integration variable as $-E= \varepsilon + \hbar \omega - E_{\bm k\nu}$ for the first term, and $E = \varepsilon - E_{\bm k\nu'}$ for the second term.
At low temperatures ($k_{\rm B}T \ll \varepsilon_{\rm F}$), we can use the approximation $E_{\bm k\nu}-E_{\bm k\nu'} \simeq \Delta_{\rm R} (\nu - \nu')$ near the Fermi surface.
Then, the remaining integration over $E_{\bm k\nu}$ gives
\begin{align}
    &\int dE_{\bm k\nu} \Bigl[ f(-E + E_{\bm k\nu}) - f(E+ E_{\bm k\nu'}) \Bigr] \notag \\
    &= 2E - (\nu -\nu')\Delta_{\rm R},
\end{align}
leading to 
\begin{align}
    & \Pi_{\nu \nu'}^R(\omega) \notag \\
    &= -\int^\infty_{-\infty} \frac{dE}{2\pi i} \frac{1}{E-i\gamma} \frac{2E - (\nu -\nu')\Delta _R}{E+\hbar \omega - (\nu -\nu')\Delta _R + i\gamma}.
\end{align}
By evaluating the remaining energy integral, we obtain
\begin{align}
    \Pi_{\nu \nu'}^R(\omega) = \frac{\hbar \omega}{\hbar \omega - (\nu - \nu')\Delta _R + 2i\gamma}.
    \label{app:surf}
\end{align}
Substituting this result into Eq.~\eqref{app:chifspre}, the spin susceptibility is finally obtained as
\begin{align}
    \chi^{R,{\rm FS}}_{\bm 0}(\omega) &= \frac{D(\varepsilon_{\rm F})}{8} \sum_{\nu \nu'} 
        \qty(
        \frac{1}{1-\tilde \Gamma_{\parallel}^R(\omega)} 
    + \frac{ 1-\nu\nu' }{1-\tilde \Gamma_{\perp}^R(\omega)}
        ) \notag \\
        & \times 
        \frac{\hbar \omega}{\hbar \omega - (\nu-\nu')\Delta_R + 2i\gamma}
    .
\end{align}

\subsection{FMR frequency shift and damping modulation}
\label{appsec:FMR}

\begin{figure}
    \centering
    \includegraphics[width=65mm]{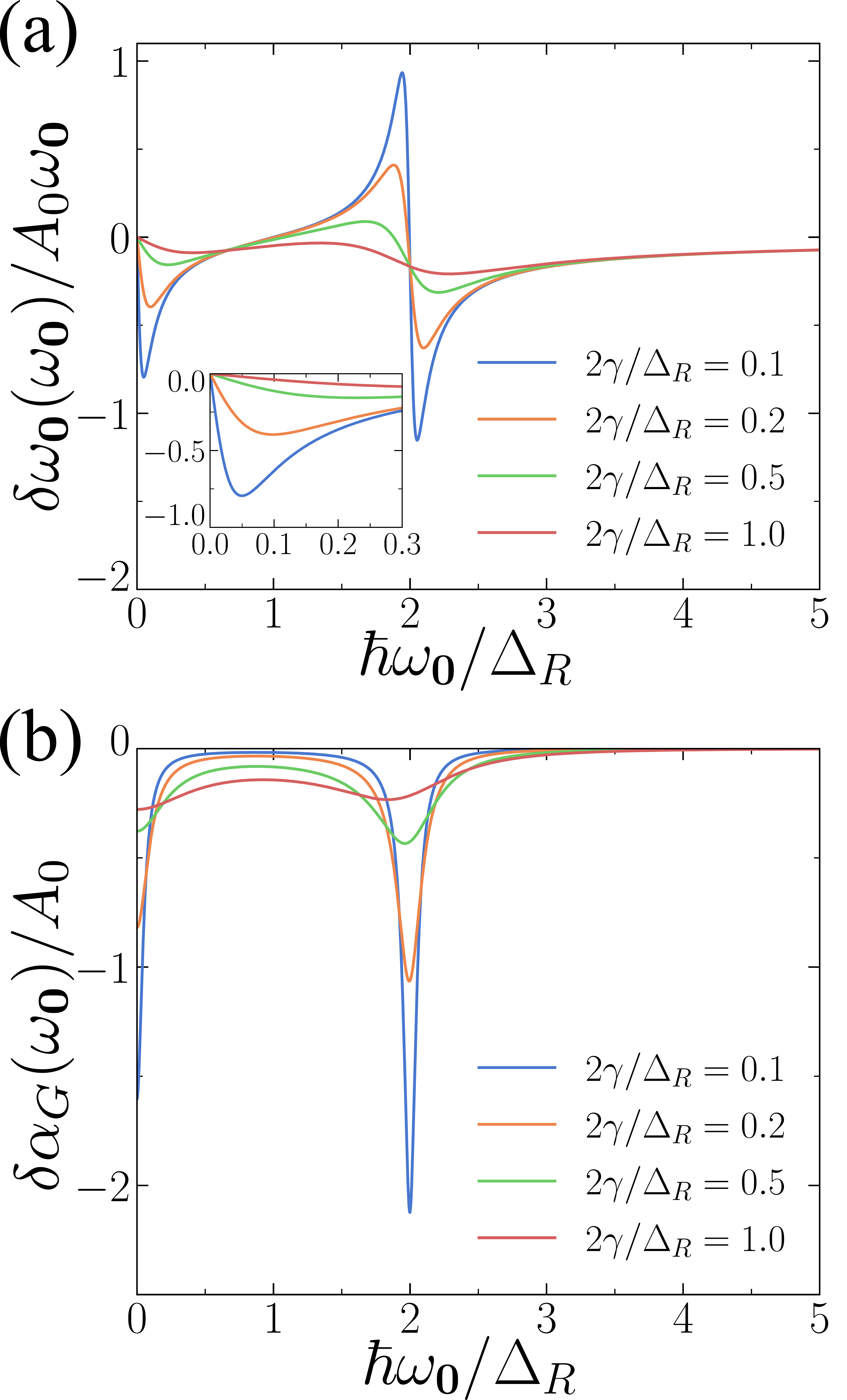}
    \caption{Resonance frequency dependence of (a) the resonance-frequency shift and (b) the damping modulation for the different values of the normalized damping rate $2\gamma/\Delta_R$, as indicated by the line colors. 
    The inset in panel (a) shows an enlarged view of the low-frequency region.}
    \label{fig:chi_plot}
\end{figure}

Combining these results with Eqs.~\eqref{ReTildeChi} and \eqref{ImTildeChi},  the FMR frequency shift and damping modulation can be evaluated.
For reference, we show the results of the FMR frequency shift and damping modulation for 2DEG with the Rashba SOC. Their frequency dependences are essentially the same as those obtained in Ref.~\cite{yama2023}, while their signs are reversed.
Figure~\ref{fig:chi_plot}(a) and (b) show the FMR shift $\delta \omega_{\bm 0}$ and the damping modulation $\delta \alpha_{\rm G}$ as functions of the resonant frequency $\omega_{\bm 0}$, respectively.
We observe dispersive and resonant behaviors for $\delta\omega_{\bm 0}$ and $\delta\alpha_{\rm G}$, respectively, at $\hbar \omega_{\bm 0} = 2\Delta_{\rm R}$ associated with the spin excitation, whose energy matches that of vertical transitions between the two spin-splitting bands in 2DEG.
In the dc limit, $\delta\omega_{\bm 0}$ approaches zero, while $\delta\alpha_{\rm G}$ remains finite due to elastic spin flipping, which is induced by the effective transverse Zeeman field through the interfacial exchange coupling.
These features demonstrate that the FMR frequency shift and damping modulation provide complementary probes of the dispersive and dissipative spin dynamics of the adjacent Rashba system~\cite{yama2023}.

\subsection{Rashba--Edelstein conductivity}
\label{app:REEconductivity}

We first recall that the Rashba--Edelstein conductivity is expressed within linear-response theory as
\begin{align}
    \sigma_{\rm REE}(\omega)
    =
    \frac{K^R_{yx}(\omega)-K^R_{yx}(0)}{i\omega},
    \label{eq:REE_conductivity}
\end{align}
where $K^R_{yx}(\omega)$ is the retarded correlation function between the spin density and the charge current.
\begin{align}
K^R_{yx}(\omega) &= \frac{i}{\hbar \mathcal A} \int^\infty_0 dt e^{i(\omega + i\delta) t} \ave{ \qty[ \hat s^y(t), \hat j_{e,x} ] }_{\mathcal H_e} ,
    \label{eq:correlation} \\
    \hat j_{e,x} &= -\frac{e}{\mathcal A} \sum_{\bm k \sigma \sigma'} c^\dagger_{{\bm k}\sigma} (v_x)_{\sigma \sigma'} c_{{\bm k} \sigma'},
\end{align}
where $\hat j_{e,x}$ is the charge-current density operator, and $v_x = (\hbar k_x/m) \hat{I} -  (\alpha_R/\hbar) \sigma^y$ is the bare velocity vertex.
The Fermi-surface contribution $\sigma_{\rm REE}^{\rm FS}(\omega)$ is obtained from Eq.~\eqref{eq:REE_conductivity} by replacing
$K^R_{yx}(\omega)$ with its Fermi-surface contribution
$K_{yx}^{R,\rm FS}(\omega)$, which is obtained by analytic continuation of the Matsubara correlation function
\begin{align}
    & K_{yx}^{\rm FS}(i\omega_\lambda) \notag \\
    &= \frac{e}{2\beta \mathcal A}
    \sum_{\bm k,n}
    \Theta(\varepsilon_n, \omega_\lambda)
    \Tr \qty[
    \Gamma^y
    g_{\bm k}(i\varepsilon_n+i\omega_\lambda)
    v_x
    g_{\bm k}(i\varepsilon_n)
    ]. \label{eq:Kyxformula}
\end{align}
Here, $\Gamma^y$ is the $y$ component of the vertex function.
By replacing the summation over the momentum, $K_{yx}^{\rm FS}(i\omega_\lambda)$ is calculated as
\begin{align}
    K_{yx}^{\rm FS}(i\omega_\lambda) &= \frac{e}{2\beta} \frac{1}{1-\tilde \Gamma_{\parallel}(i\omega_\lambda)}
    \sum_{\nu \nu'} \int \frac{dk \, k}{2\pi} \int \frac{d\varphi}{2\pi} \notag \\
    & \hspace{-5mm} \times \sum_n \Theta(\varepsilon_n, \omega_\lambda)  \frac{\Tr [\sigma^y \mathcal P_{\nu} \qty(\dfrac{\hbar k_x}{m} - \dfrac{\alpha_R}{\hbar} \sigma^y ) \mathcal P_{\nu'} ]}{\mathcal D_{\bm k\nu}(i\varepsilon_n + i\omega_\lambda) \mathcal D_{\bm k\nu'}(i\varepsilon_n)} .
\end{align}
The integral over the angle $\varphi$ is calculated as
\begin{align}
    & \int^{2\pi}_0 \frac{d\varphi}{2\pi} \Tr [\sigma^y \mathcal P_{\nu} \qty(\frac{\hbar k_x}{m} - \frac{\alpha_R}{\hbar} \sigma^y ) \mathcal P_{\nu'} ] \notag \\
    & \quad = -\frac{1}{2} \qty( \delta_{\nu \nu'} \nu \frac{\hbar k}{m} + \frac{\alpha_R}{\hbar} )
\end{align}
Using Eq.~(\ref{app:Pi-def}), the response function can be expressed as
\begin{align}
    K_{yx}^{\rm FS}(i\omega_\lambda)  &= -\frac{e}{4}  
    \frac{1}{1-\tilde{\Gamma}_{\parallel}(i\omega_\lambda)}
    \sum_{\nu \nu'} D_\nu(\varepsilon_{\rm F}) \notag \\
    &\times \qty( \delta_{\nu \nu'} \nu \frac{\hbar k_{F\nu}}{m} + \frac{\alpha_R}{\hbar} ) \Pi_{\nu \nu'}(i\omega_\lambda),
\end{align}
where $k_{{\rm F}\nu} = k_{\rm F}(1-\nu \Delta_{\rm R}/2\varepsilon_{\rm F})$ is the Fermi wavenumber of the band $\nu$.
Performing the analytic continuation $i\omega_\lambda \to \hbar \omega +i\delta$, the retarded component of the correlation function is given by
\begin{align}
    K_{yx}^{R,\rm FS}(\omega)  &= -\frac{e}{4}   
    \sum_{\nu \nu'} D_\nu(\varepsilon_{\rm F}) \qty( \delta_{\nu \nu'} \nu \frac{\hbar k_{F\nu}}{m} + \frac{\alpha_R}{\hbar} )  \notag \\
    & \hspace{15mm} \times \frac{\Pi^R_{\nu \nu'}(\omega)}{1-\tilde{\Gamma}^R_{\parallel}(\omega)}  .
\end{align}
By combining this result with Eqs.~\eqref{app:surf} and Eq.~\eqref{eq:REE_conductivity}, we obtain Eq.~\eqref{eq:sigmaREEformula} in the main text.
 
\bibliography{references}
    
\end{document}